\documentclass[twocolumn,10pt]{IEEEtran}

\usepackage{amsmath}
\usepackage{mathtools}
\usepackage{enumerate}
\usepackage{paralist}
\usepackage{graphicx}
\usepackage{epstopdf}
\graphicspath{{Figures/}}
\usepackage{subfigure}
\usepackage{cite, color}
\usepackage{color}
\usepackage{array}
\usepackage{booktabs}
\usepackage{comment}
\usepackage{cases}
\usepackage{algorithm}
\usepackage{algorithmic}
\usepackage{mathrsfs} 
\usepackage{url}
\usepackage{amssymb}
\usepackage{amsthm} 

\newtheorem{theorem}{Theorem}

\newtheorem{proposition}[theorem]{Proposition}

\newtheorem{statement}{Statement}

\begin{document}

\title{Resource Allocation Under Channel Uncertainties for Relay-Aided Device-to-Device Communication Underlaying LTE-A Cellular  Networks}

\author{Monowar Hasan, Ekram Hossain, and Dong In Kim
\thanks{M. Hasan and E. Hossain are with the Department of Electrical and Computer Engineering, University of Manitoba, Winnipeg, Canada (emails:  monowar\_hasan@umanitoba.ca, Ekram.Hossain@umanitoba.ca).  
D. I. Kim is with the School of Information and Communication
Engineering at the Sungkyunkwan University (SKKU), Korea (email: dikim@skku.ac.kr).} 

}

\maketitle

\begin{abstract}
Device-to-device (D2D) communication in cellular networks allows direct transmission between two cellular devices with local communication needs.
Due to the increasing number of autonomous heterogeneous devices in future mobile networks, an efficient resource allocation scheme is required to maximize network throughput and achieve higher spectral efficiency.
In this paper, performance of network-integrated D2D communication under channel uncertainties is investigated where D2D traffic is carried through relay nodes. Considering a multi-user and multi-relay network, we propose a robust distributed solution for resource allocation with a view to maximizing network sum-rate when the interference from other relay nodes and the link gains are uncertain. An optimization problem is formulated for allocating radio resources at the relays to maximize end-to-end rate as well as satisfy the quality-of-service (QoS) requirements for cellular and D2D user equipments under total power constraint. Each of the uncertain parameters is modeled by a bounded distance between its estimated and bounded values. We show that  the robust problem is convex and a gradient-aided dual decomposition algorithm is applied to allocate radio resources in a distributed manner. Finally, to reduce the cost of robustness defined as the reduction of achievable sum-rate, we utilize the \textit{chance constraint approach} to achieve a trade-off between robustness and optimality. The numerical results show that there is a distance threshold beyond which relay-aided D2D communication significantly improves network performance when compared to direct communication between D2D peers.

\end{abstract}

\begin{IEEEkeywords}
D2D communication, LTE-Advanced  Layer 3 (L3) relay, robust worst-case resource allocation,  uncertain channel gain, ellipsoidal uncertainty set, chance constraint.
\end{IEEEkeywords}

\section{Introduction} \label{sec:intro}

Device-to-device (D2D) communication enables wireless peer-to-peer services directly between user equipments (UEs) to facilitate high data rate local service as well as offload the traffic of cellular base station (i.e., Evolved Node B [eNB] in  an LTE-Advanced [LTE-A] network). By reusing the LTE-A cellular resources, D2D communication enhances spectrum utilization and improves cellular coverage. In conjunction with traditional local voice and data services, D2D communication opens up new opportunities for commercial applications, such as proximity-based services, in particular social networking applications with content sharing features (i.e., exchanging photos, videos or documents through smart phones), local advertisement, multi-player gaming and data flooding \cite{d2d_example, d2d_example2, 3gpp:d2d_example}.

In the context of D2D communication, it becomes a crucial issue to set up direct links between the D2D UEs while satisfying the quality-of-service (QoS) requirements of traditional cellular UEs (CUEs) and the D2D UEs in the network.  In practice, the advantages of D2D communication may be limited due to: 
\begin{inparaenum}[\itshape i)]
\item \textit{longer distance:} the potential D2D UEs may not be in near proximity;
\item \textit{poor propagation medium:} the link condition between two D2D UEs may not be favourable;
\item \textit{interference to and from CUEs:} in a spectrum underlay system, D2D transmitters can cause severe interference to other receiving nodes in the network and also the D2D receivers may experience interference from other transmitting nodes. Partitioning the available spectrum for its use by CUEs and D2D UEs in a non-overlapping manner (i.e., overlay D2D communication) could be an alternative; however, this would significantly reduce spectrum utilization \cite{phond-d2d, d2d_swarm}. 
\end{inparaenum}
In such cases, network-assisted transmission through relays could enhance the performance of D2D communication when D2D UEs are far away from each other and/or the quality of D2D communication channel is not good enough for direct communication.

Unlike most of the existing work on D2D communication, in this paper, we consider relay-assisted D2D communication in LTE-A cellular networks where D2D pairs are served by the relay nodes. In particular, we consider LTE-A Layer-3 (L3) relays\footnote{An L3 relay with self-backhauling configuration performs the same operation as an eNB except that it has a  lower transmit power and a smaller cell size. It controls cell(s) and each cell has its own cell identity. The relay transmits its own control signals and the UEs are able to receive scheduling information directly from the relay node \cite{relay-book-1}.}. We concentrate on scenarios in which the proximity and link condition between the potential D2D UEs  may not be favorable for direct communication.  Therefore, they may communicate via relays. The radio resources at the relays (e.g., resource blocks [RBs] and transmission power) are shared among the D2D communication links and the two-hop cellular links using these relays. An use-case for such relay-aided D2D communication could be the machine-to-machine (M2M) communication \cite{d2d_m2m_1} for smart cities. In such a communication scenario,  automated sensors (i.e., UEs) are deployed within a macro-cell ranging a few city blocks; however, the link condition and/or proximity between devices may not be favorable. Due to the nature of applications, these UEs are required to periodically transmit data \cite{m2m_our_paper}. Relay-aided D2D communication could be an elegant solution to provide reliable transmission as well as improve overall network throughput in such a scenario.   

Due to time-varying and random nature of wireless channel, we formulate a robust resource allocation problem with an objective to maximizing the end-to-end rate (i.e., minimum achievable rate over two hops) for the UEs while maintaining the QoS (i.e., rate) requirements for cellular and D2D UEs under total power constraint at the relay node. The link gains, the interference among relay nodes and interference at the receiving D2D UEs are not exactly known (i.e., estimated with an additive error). The robust problem formulation is observed to be convex, and therefore, we apply a gradient-based method to solve the problem distributively at each relay node with polynomial complexity. We demonstrate that introducing robustness to deal with channel uncertainties affects the achievable network sum-rate. To reduce the cost of robustness defined as the corresponding reduction of achievable sum-rate, we utilize the \textit{chance constraint approach} to achieve a trade-off between robustness and optimality by adjusting some protection functions. We compare the performance of our proposed method with an underlay D2D communication scheme where the D2D UEs communicate directly without the assistance of relays. The numerical results show that after a distance threshold for the D2D UEs, relaying D2D traffic provides significant gain in achievable data rate. The main contributions of this paper can be summarized as follows:

\begin{itemize}
\item We analyze the performance of relay-assisted D2D communication under uncertain system parameters. The problem of RB and power allocation at the relay nodes for the CUEs and D2D UEs is formulated and solved for the globally optimal solution when perfect channel gain information for the different links is available.  As opposed to  most of the resource allocation schemes in the literature where only a single D2D link is considered, we consider multiple D2D links along with multiple cellular links that are supported by relay nodes.

\item Assuming that the perfect channel information is unavailable, we formulate a robust resource allocation problem for relay-assisted D2D communication under uncertain channel information in both the hops and show that the convexity of the robust formulation is maintained.  We propose a distributed algorithm with a polynomial time complexity.

\item The cost of robust resource allocation is analyzed. In order to achieve a balance between the network performance and robustness, we provide a trade-off mechanism. 
\end{itemize}

The rest of this paper is organized as follows. A review of the related work and motivation of this work are presented in Section \ref{sec:related_works}. In Section \ref{sec:sys_model}, we present the system model and assumptions. In Section~\ref{sec:nominal}, we formulate the RB and power allocation problem for the nominal (i.e., non-robust) case. The robust resource allocation problem is formulated in Section~\ref{sec:robust}. In order to allocate resources efficiently, we propose a robust distributed algorithm and discuss the robustness-optimality trade-off in Section \ref{sec:robust_algo}. The performance evaluation results are presented in Section \ref{sec:performance_eval} and finally we conclude the paper in Section \ref{sec:conclusion}. The key mathematical notations used in the paper are listed in Table \ref{tab:notations}.

\begin{table}[!t]
\renewcommand{\arraystretch}{1.3}
\caption{Mathematical Notations}
\label{tab:notations}
\centering
\begin{tabular}{c|p{5.5cm}}
\hline
\bfseries Notation & \bfseries Physical interpretation\\
\hline\hline
$\mathcal{N} = \lbrace 1, 2, \ldots, N \rbrace $ & Set of available RBs \\
\hline $\mathcal{L} = \lbrace 1, 2, \ldots, L \rbrace$ & Set of relays \\
\hline $u_l$  & A UE served by relay $l$ \\
\hline $\mathcal{U}_l, |\mathcal{U}_l|$ & Set of UEs and total number of UEs served by relay $l$, respectively \\
\hline $h_{i,j}^{(n)}$ & Direct link gain between the node $i$ and $j$ over RB $n$ \\
\hline $R_{u_l}^{(n)}$ & End-to-end data rate for $u_l$ over RB $n$ \\
\hline $x_{u_l}^{(n)}, S_{u_l, l}^{(n)}$ & RB allocation indicator and actual transmit power for $u_l$ over RB $n$, respectively \\
\hline $I_{u_l, l}^{(n)}$ & Aggregated interference experienced by $u_l$ over RB $n$ \\
\hline $\mathbf{g}_{l, i}^{(n)}$ & Nominal link gain vector over RB $n$ in hop $i$ \\
\hline $\bar{\mathbf{g}}_{l, i}^{(n)}$, $\hat{\mathbf{g}}_{l, i}^{(n)}$ & Estimated and uncertain (i.e., the bounded error) link gain vector, respectively, over RB $n$ in hop $i$ \\
\hline $\Re_{g_{l, i}}^{(n)},  \Delta_{g_{l, i}}^{(n)} $ & Uncertainty set and protection function, respectively, for link gain over RB $n$ in hop $i$ \\
\hline $\Re_{I_{u_l,l}}^{(n)},  \Delta_{I_{u_l,l}}^{(n)}$ & Uncertainty set  and protection function of interference level, respectively, for $u_l$ over RB $n$ \\
\hline $\Psi_{l,i}^{(n)}$ & Bound of uncertainty in link gain for hop $i$ over RB $n$ \\
\hline $\Upsilon_{u_l}^{(n)} $ & Bound of uncertainty in interference level for $u_l$ over RB $n$ \\
\hline ${\parallel \mathbf{y} \parallel}_\alpha$ & Linear norm of vector $\mathbf{y}$ with order $\alpha$ \\
\hline ${\parallel \mathbf{y} \parallel}^*$ & Dual norm of $\parallel \mathbf{y} \parallel$\\
\hline $\mathbf{abs}\lbrace y \rbrace$ & Absolute value of $y$ \\
\hline $\mathbf{A}(j,:)$ & $j$-th row of matrix $\mathbf{A}$ \\
\hline $\Lambda_{\kappa}^{(t)}$ & Step size for variable $\kappa$ at iteration $t$\\
\hline $\mathscr{R}_\Delta $ & Reduction of achievable sum-rate due to uncertainty\\
\hline $\Theta_{l,i}^{(n)}$ & Threshold probability of violating interference constraint for RB $n$ in hop $i$ \\
\hline $\mathcal{S}_{\Theta_{l,i}^{(n)}} \left( \mathscr{R}_\Delta \right)$ & Sensitivity of $\mathscr{R}_\Delta $ in hop $i$ over RB $n$ \\
\hline 
\end{tabular}
\end{table}

\section{Related Work and Motivation} \label{sec:related_works}

Although resource allocation for D2D communication in future generation orthogonal frequency-division multiple access (OFDMA)-based wireless networks is one of the active areas of research, there are very few work which consider relays for D2D communication. A resource allocation scheme based on a column generation method is proposed in \cite{phond-d2d} to maximize the spectrum utilization by finding the minimum transmission length (i.e., time slots) for D2D links while protecting the cellular users from interference and guaranteeing QoS. In \cite{zul-d2d}, a greedy heuristic-based resource allocation scheme is proposed for both uplink and downlink scenarios where a D2D pair shares the same resources with CUE only if the achieved SINR is greater than a given SINR requirement. A new spectrum sharing protocol for D2D communication overlaying a cellular network is proposed in \cite{d2d_new_paper}, which allows the D2D users to communicate bi-directionally while assisting the two-way communications between the eNB and the CUE. The resource allocation problem for D2D communication underlaying cellular networks is addressed in \cite{lingyang-icc13}.  In \cite{xen-1}, the authors consider relay selection and resource allocation for uplink transmission in LTE-Advanced (LTE-A) networks with two classes of users having different (i.e., specific and flexible) rate requirements. The objective is to maximize system throughput by satisfying the rate requirements for the rate-constrained users while confining the transmit power within a power-budget.

Although D2D communication was initially proposed to relay user traffic \cite{d2d_first_relay}, not many work consider using relays in D2D communication. To the best of our knowledge, relay-assisted D2D communication was first introduced in \cite{d2d-rel-4} where the relay selection problem for D2D communication underlaying cellular network was studied. The authors propose a distributed relay selection method for relay assisted D2D communication system which firstly coordinates the interference caused by the coexistence of D2D system and cellular network and eliminates improper relays correspondingly. Afterwards, the best relay is chosen among the optional relays using a distributed method. In \cite{d2d-rel-3}, the authors consider D2D communication for relaying  UE traffic toward the eNB and deduce a relay selection rule based on the interference constraints. In \cite{d2d-rel-1,  d2d_relay_2}, the maximum ergodic capacity and outage probability of cooperative relaying are investigated in relay-assisted D2D communication  considering power constraints at the eNB. The numerical results show that multi-hop relaying lowers the outage probability and improves cell edge throughput capacity by reducing the effect of interference from the CUE.  


In all of the above cited work, it has generally been assumed that complete system information (e.g., channel state information [CSI]) is available to the network nodes, which is unrealistic for a practical system. 
Uncertainty in the CSI (in particular the channel quality indicator [CQI] in an LTE-A system) can be modeled by sum of estimated CSI (i.e., the nominal value) and some additive error (the uncertain element). Accordingly, by using robust optimization theory, the nominal optimization problem (i.e., the optimization problem without considering uncertainty) is mapped to another optimization problem (i.e., the robust problem). To tackle uncertainty, two approaches have commonly been used in robust optimization theory. First, the \textit{Bayesian approach} (Chapter 6.4 in \cite{book-boyd}) considers the statistical knowledge of errors and satisfies the optimization constraints in a probabilistic manner. Second, the \textit{worst-case approach} (Chapter 6.4 in \cite{book-boyd}, \cite{worst-case_robust}) assumes that the error (i.e., uncertainty) is bounded in a closed set called the \textit{uncertainty set} and satisfies the constraints for all realizations of the uncertainty in that set.  Although the Bayesian approach has been widely used in the literature (e.g., in \cite{bayesian_robust_1}, \cite{bayesian_robust_2}), the worst-case approach is more appropriate due to the fact that it satisfies the constraints in all error instances. By applying the worst-case approach, the size of the uncertainty set can be obtained from the statistics of error. As an example, the uncertainty set can be defined by a probability distribution function of uncertainty in such a way  that  all realizations of uncertainty remain within the uncertainty set with a given probability.

Applying robustness brings in new variables in the optimization problem, which may change the nominal formulation to a non-convex optimization problem and require excessive calculations to solve. To avoid this difficulty,  the robust problem is converted to a convex optimization problem and solved in a traditional way. Although not in the context of D2D communication, there have been a few work considering resource allocation under uncertainties in the radio links. One of the first contributions dealing with channel uncertainties is \cite{uncertainity-early} where the author models the time-varying communication for single-access and multiple-access channels without feedback. For an OFDMA system, the resource allocation problem under channel uncertainty for a cognitive radio (CR) base station communicating with multiple CR mobile stations is considered in \cite{uncertainity-cr-2} for downlink communication. Two robust power control schemes are developed in \cite{robust_rel} for a CR network  with cooperative relays. In \cite{robust_pow}, a  robust power control algorithm is proposed for a CR network to maximize the social utility defined as the network sum rate. A robust worst-case interference control mechanism is provided in \cite{robust_intf} to maximize rate while keeping the interference to primary user below a threshold. 

Taking the advantage of L3 relays supported by the 3GPP standard, in our earlier work \cite{d2d_our_paper}, we studied the performance of network-assisted D2D communications assuming the availability of perfect CSI and showed that relay-aided D2D communication provides significant performance gain for long distance D2D links. In this paper, we extend the work utilizing the theory of worst-case robust optimization to maximize the end-to-end data rate under link uncertainties for the UEs with minimum QoS requirements while protecting the other receiving relay nodes and D2D UEs from interference. To make the robust formulation more tractable and obtain a near-optimal solution for satisfying all the constraints in the nominal problem, we apply the notion of \textit{protection function} instead of uncertainty set. 

\section{System Model and Assumptions} \label{sec:sys_model}

\subsection{Network Model}

\begin{figure}[!t]
\centering
\includegraphics[scale=0.65]{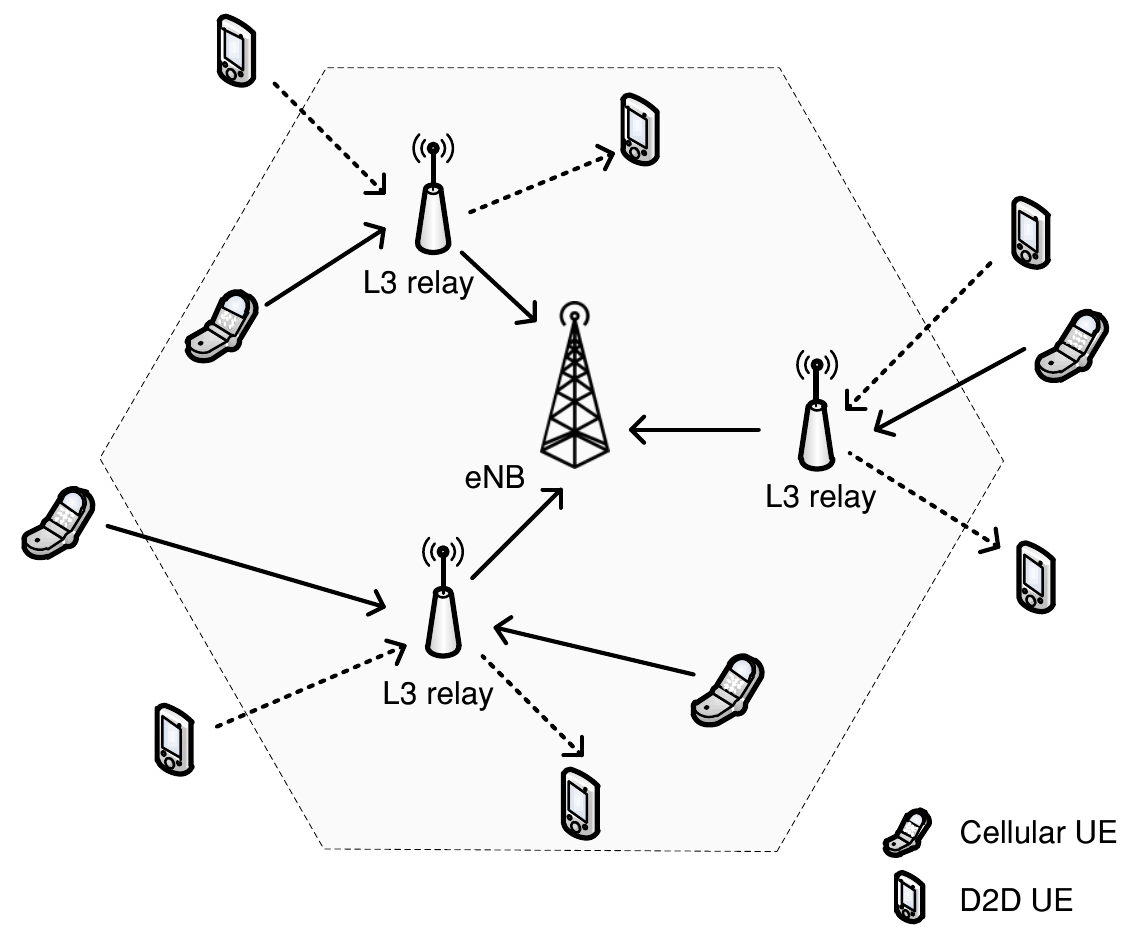}
\caption{A single cell with multiple relay nodes. We assume that the CUE-eNB links are unfavourable for direct communication and they need the assistance of relays. The D2D UEs are also supported by the relay nodes due to long distance and/or poor link condition between peers. 
} 
\label{fig:nw_diagram}
\end{figure}

A relay node in LTE-A is connected to the radio access network (RAN) through a donor eNB with a wireless connection and serves  both the cellular and D2D UEs. Let $\mathcal{L} = \lbrace 1, 2, \ldots, L \rbrace$ denote the set of fixed-location Layer 3 (L3) relays in the network as shown in Fig. \ref{fig:nw_diagram}. The system bandwidth is divided into $N$ RBs denoted by $\mathcal{N} = \lbrace 1, 2, \ldots, N \rbrace$ which are used by all the relays in a spectrum underlay fashion. When the link condition between two D2D peers is too poor for direct communication,  scheduling and resource allocation for the D2D UEs can be done in a relay node (i.e., L3 relay) and the D2D traffic can be transmitted through that relay. We refer to this as \textit{relay-aided D2D communication} which can be an efficient approach to provide better quality-of-service (QoS) for communication between distant D2D UEs.

The CUEs and D2D pairs  constitute set $\mathcal{C} = \lbrace 1, 2, \ldots, C \rbrace$ and $\mathcal{D} = \lbrace 1, 2, \ldots, D \rbrace$, respectively, where the D2D pairs are discovered during the D2D session setup. We assume that the CUEs are outside the coverage region of eNB and/or having bad channel condition, and therefore, the CUE-eNB communications need to be supported by the relays. Besides, direct communication between two D2D UEs requires the assistance of a relay node. We assume that association of the UEs (both cellular and D2D)  to the corresponding relays are performed before resource allocation. The UEs assisted by relay $l$ are denoted by $u_l$. The set of UEs assisted by relay $l$ is $\mathcal{U}_l$ such that $\mathcal{U}_l \subseteq \lbrace \mathcal{C} \cup \mathcal{D} \rbrace, \forall l \in \mathcal{L}$, $\bigcup_l \mathcal{U}_l = \lbrace \mathcal{C} \cup \mathcal{D} \rbrace$, and  $\bigcap_l \mathcal{U}_l = \varnothing$.  


We assume that during a certain time instance,  only one relay-eNB link is active in the second hop to carry CUEs' data (i.e., transmissions of relays to the eNB in the second hop are orthogonal in time). Scheduling of the relays for transmission in the second hop is done by the eNB.\footnote{Scheduling of relay nodes by the eNB is out of the scope of this paper.} However, multiple relays can transmit to their corresponding D2D UEs in the second hop. Note that, in the first hop, the transmission between a UE (i.e., either CUE or D2D UE) and relay can be considered an uplink communication. In second hop,  the transmission between a relay and the eNB can be considered an uplink communication from the perspective of the eNB  whereas the transmission from a relay to a D2D UE can be considered as a downlink communication. In our system model, taking advantage of the capabilities of L3 relays, scheduling and resource allocation for the UEs is performed in the relay nodes to reduce the computation load at the eNB.

\subsection{Achievable Data Rate}

We denote by $h_{i,j}^{(n)}$ the direct link gain between node $i$ and $j$ over RB $n$. The interference link gain between relay (UE) $i$  and UE (relay) $j$ over RB $n$ is denoted by $g_{i,j}^{(n)}$  where UE (relay) $j$ is not associated with relay (UE) $i$.  The unit power ${\rm SINR}$ for the link between UE $u_l \in \mathcal{U}_l$ and relay $l$ using RB $n$ in the first hop is given by
\begin{equation}
\label{eq:SINR_1}
\gamma_{u_l, l, 1}^{(n)} = \frac{h_{u_l, l}^{(n)}}{\displaystyle \sum_{\forall u_j \in \mathcal U_j, j \neq l, j \in \mathcal{L}} P_{u_j, j}^{(n)} g_{u_j, l}^{(n)} + \sigma^2}.
\end{equation}
The unit power ${\rm SINR}$ for the link between relay $l$ and eNB  for CUE (i.e.,  $u_l \in \lbrace \mathcal{C} \cap \mathcal{U}_l \rbrace$) in the second hop is as follows: 
\begin{equation}
\label{eq:SINR_2}
\gamma_{l, u_l, 2}^{(n)} = \frac{h_{l, eNB}^{(n)}}{\displaystyle \sum_{\forall u_j \in \lbrace \mathcal{D} \cap \mathcal{U}_j \rbrace, j \neq l, j \in \mathcal{L}} P_{j, u_j}^{(n)} g_{j, eNB}^{(n)} + \sigma^2}.
\end{equation}

Similarly, the unit power ${\rm SINR}$ for the link between relay $l$ and receiving D2D UE for the D2D-pair (i.e., $u_l \in \lbrace \mathcal{D} \cap \mathcal{U}_l \rbrace$) in the second hop can be written as
\begin{equation}
\label{eq:SINR_3}
\gamma_{l, u_l, 2}^{(n)} = \frac{h_{l, u_l}^{(n)}}{\displaystyle \sum_{\forall u_j \in  \mathcal{U}_j , j \neq l, j \in \mathcal{L}} P_{j, u_j}^{(n)} g_{j, u_l}^{(n)} + \sigma^2}.
\end{equation}

In  (\ref{eq:SINR_1})--(\ref{eq:SINR_3}), $P_{i, j}^{(n)}$ is the transmit power in the  link between $i$ and $j$ over RB $n$, $\sigma^2 =  N_0 B_{RB}$ where $B_{RB}$ is bandwidth of an RB, and $N_0$ denotes thermal noise.  $h_{l, eNB}^{(n)}$ is the gain in the relay-eNB link and $h_{l, u_l}^{(n)}$ is the gain in the link between relay $l$ and receiving D2D UE corresponding to the D2D transmitter UE $u_l$.

The achievable data rate for $u_l$ in the first hop can be expressed as $r_{u_l, 1}^{(n)} = B_{RB} \log_2 \left( 1 + P_{u_l, l}^{(n)} \gamma_{u_l, l, 1}^{(n)} \right)$. Note that, this rate expression is valid under the assumption of Gaussian (and spectrally white) interference which holds for a large number of interferers. Similarly, the achievable data rate in the second hop is $r_{u_l, 2}^{(n)} = 
B_{RB} \log_2 \left( 1 + P_{l, u_l}^{(n)} \gamma_{l, u_l, 2}^{(n)} \right)$. Since we are considering a two-hop communication, the end-to-end data rate for  $u_l$ on RB $n$ is half of the minimum achievable data rate over two hops \cite{mutihop-rate}, i.e.,   
\begin{equation}
\label{eqn:e2e_rate}
R_{u_l}^{(n)} =  \frac{1}{2} \min \left\lbrace  r_{u_l, 1}^{(n)} , r_{u_l, 2}^{(n)}    \right\rbrace.
\end{equation}

\section{Resource Block (RB) and Power Allocation in Relay Nodes}
\label{sec:nominal}

\subsection{Formulation of the Nominal Resource Allocation Problem}

For each relay, the objective of radio resource (i.e., RB and transmit power) allocation is to obtain  the assignment of RB and power level to the UEs that maximizes the system capacity, which is defined as the minimum achievable data rate over two hops. Let the maximum allowable transmit power for UE (relay) is $P_{u_l}^{max}$ ($P_l^{max}$). The RB allocation indicator is a binary decision variable $x_{u_l}^{(n)} \in \lbrace 0, 1\rbrace$, where 
\begin{equation}
x_{u_l}^{(n)}  = \begin{cases}
 1, \quad  \text{if RB $n$ is assigned to UE $u_l$} \\
 0, \quad \text{otherwise.}
\end{cases}
\end{equation}
Let $\displaystyle R_{u_l} = \sum_{n =1}^N  x_{u_l}^{(n)} R_{u_l}^{(n)}$ denotes the achievable sum-rate over allocated RB(s) and let the QoS (i.e., rate) requirements for UE $u_l$ is denoted by $Q_{u_l}$. Considering that the same RB(s) will be used by the relay in both the hops (i.e., for communication between relay and eNB and between relay and D2D UEs), the resource allocation problem for each relay $l \in \mathcal{L}$ can be stated as follows:
\begin{subequations}
\setlength{\arraycolsep}{0.0em}
\begin{eqnarray}
\mathbf{(P1)} ~ \underset{x_{u_l}^{(n)}, P_{u_l, l}^{(n)}, P_{l, u_l}^{(n)}}{\operatorname{max}} ~ \sum_{u_l \in \mathcal{U}_l } &&  \sum_{n =1}^N   x_{u_l}^{(n)} R_{u_l}^{(n)}  \nonumber \\
\text{subject to} \quad \sum_{u_l \in \mathcal{U}_l} x_{u_l}^{(n)} ~ &&{\leq} ~ 1,  \quad ~~~\forall n \in \mathcal{N} \label{eq:con-bin} \\
\quad \sum_{n =1}^N x_{u_l}^{(n)} P_{u_l, l}^{(n)} ~ &&{\leq} ~ P_{u_l}^{max}, ~ \forall u_l \in \mathcal{U}_l \label{eq:con-pow-ue} \\
\quad \sum_{u_l \in \mathcal{U}_l } \sum_{n =1}^N x_{u_l}^{(n)} P_{l, u_l}^{(n)} ~ &&{\leq} ~ P_l^{max}  \label{eq:con-pow-rel} \\
\quad \sum_{u_l \in \mathcal{U}_l } x_{u_l}^{(n)} P_{u_l, l}^{(n)} g_{{u_l^*}, l, 1}^{(n)} ~ &&{\leq} ~ I_{th, 1}^{(n)},  ~~\forall n \in \mathcal{N} \label{eq:con-intf-1}\\
\quad \sum_{u_l \in \mathcal{U}_l } x_{u_l}^{(n)}  P_{l, u_l}^{(n)} g_{l, {u_l^*}, 2}^{(n)} ~ &&{\leq} ~ I_{th, 2}^{(n)}, ~~\forall n \in \mathcal{N} \label{eq:con-intf-2} \\
\quad R_{u_l}  ~ &&{\geq} ~ Q_{u_l}, \quad \forall u_l \in \mathcal{U}_l  \label{eq:con-QoS-cue}\\
\quad P_{u_l, l}^{(n)} ~ \geq ~ 0, ~~ P_{l, u_l}^{(n)} ~ &&{\geq} ~ 0,  \quad ~~~\forall n \in \mathcal{N}, u_l \in \mathcal{U}_l \label{eq:con-pow-0}
\end{eqnarray}
\setlength{\arraycolsep}{5pt}
\end{subequations}
where the rate of $u_l$ over RB $n$ 
$$R_{u_l}^{(n)} = \frac{1}{2} \min \left\lbrace \begin{matrix}
B_{RB} \log_2 \left( 1 + P_{u_l, l}^{(n)} \gamma_{u_l, l, 1}^{(n)} \right), \vspace*{0.5em}\\
B_{RB} \log_2 \left( 1 + P_{l, u_l}^{(n)} \gamma_{l, u_l, 2}^{(n)} \right)   
\end{matrix} \right\rbrace$$ 
and the unit power ${\rm SINR}$  for the first hop, \[\gamma_{u_l, l, 1}^{(n)} = \frac{h_{u_l, l}^{(n)}}{ I_{u_l,l,1}^{(n)} + \sigma^2  }\] and the unit power ${\rm SINR}$ for the second hop, 
\begin{numcases}{\gamma_{l, u_l, 2}^{(n)} = }
\frac{h_{l, eNB}^{(n)}}{I_{l, u_l,2}^{(n)} + \sigma^2}, &   $u_l \in \lbrace \mathcal{C} \cap \mathcal{U}_l \rbrace$ \nonumber \\
\frac{h_{l, u_l}^{(n)}}{I_{l,u_l,2}^{(n)} + \sigma^2}, &   $u_l \in \lbrace \mathcal{D} \cap \mathcal{U}_l \rbrace$. \nonumber
\end{numcases}
In the above $I_{u_l,l,1}^{(n)}$ and $I_{l, u_l,2}^{(n)}$ denote the interference received by $u_l$ over RB $n$ in the first and second hop, respectively, and are given as follows:
 $I_{u_l,l,1}^{(n)} = \displaystyle \sum_{\forall u_j \in \mathcal U_j, j \neq l, j \in \mathcal{L}}  x_{u_j}^{(n)} P_{u_j, j}^{(n)} g_{u_j, l}^{(n)}$
\begin{numcases}{I_{l,u_l,2}^{(n)} = }
\displaystyle \sum_{\forall u_j \in \lbrace \mathcal{D} \cap \mathcal{U}_j \rbrace, j \neq l, j \in \mathcal{L}} x_{u_j}^{(n)} P_{j, u_j}^{(n)} g_{j, eNB}^{(n)} , & \hspace{-2em} $u_l \in \lbrace \mathcal{C} \cap \mathcal{U}_l \rbrace$ \nonumber \\ 
\displaystyle \sum_{\forall u_j \in  \mathcal{U}_j , j \neq l, j \in \mathcal{L}} x_{u_j}^{(n)}  P_{j, u_j}^{(n)} g_{j, u_l}^{(n)}, & \hspace{-3em} $  u_l \in \lbrace \mathcal{D} \cap \mathcal{U}_l \rbrace$. \nonumber   
\end{numcases} 

With the constraint in (\ref{eq:con-bin}), each RB is assigned to only one UE. With the constraints in (\ref{eq:con-pow-ue}) and (\ref{eq:con-pow-rel}), the transmit power is limited by the maximum power budget. The constraints in (\ref{eq:con-intf-1}) and (\ref{eq:con-intf-2}) limit the amount of interference introduced to the other relays and the receiving D2D UEs in the first and second hop, respectively, to be less than some threshold. The constraint in (\ref{eq:con-QoS-cue}) ensures the minimum QoS requirements for the CUE and D2D UEs. The constraint in (\ref{eq:con-pow-0}) is the non-negativity condition for transmit power.

Similar to \cite{ref_user}, the concept of reference node is adopted here. For example, to allocate the power level considering the interference threshold in the first hop, each UE $u_l$ associated with relay node $l$  obtains the reference user $u_l^*$ associated with the other relays and the corresponding channel gain $g_{{u_l^*}, l, 1}^{(n)}$ for $\forall n$  according to the following equation:
\begin{equation}
u_{l}^* = \underset{j}{\operatorname{argmax}} ~ g_{u_l , j}^{(n)},~~ u_l \in \mathcal{U}_l, j \neq l, j \in \mathcal{L} \label{eq:ref_user1}.
\end{equation}
Similarly, in the second hop, for each relay $l$, the transmit power will be adjusted accordingly considering interference introduced to the receiving D2D UEs (associated with other relays) considering the corresponding channel gain $g_{l, {u_l^*}, 2}^{(n)}$ for $\forall n$ where the reference user  is obtained by 
\begin{equation}
u_{l}^* = \underset{u_j}{\operatorname{argmax}} ~ g_{l , u_j}^{(n)}, ~~ j \neq l, j \in \mathcal{L},  u_j \in \lbrace \mathcal{D} \cap \mathcal{U}_j \rbrace. \label{eq:ref_user2}
\end{equation}   
From (\ref{eqn:e2e_rate}), the maximum data rate for UE $u_l$ over RB $n$ is achieved when $P_{u_l, l}^{(n)} \gamma_{u_l, l, 1}^{(n)} = P_{l, u_l}^{(n)} \gamma_{l, u_l, 2}^{(n)}$. Therefore, in the second hop, the power allocated for UE $u_l$, $P_{l, u_l}$ can be expressed as a function of power allocated for transmission in the first hop, $P_{u_l, l}$ as follows: $P_{l, u_l}^{(n)}  = \frac{\gamma_{u_l, l, 1}^{(n)}}{\gamma_{l, u_l, 2}^{(n)}}P_{u_l, l}^{(n)}$. Hence the data rate for $u_l$ over RB $n$ can  be expressed as 
\begin{equation}
\label{eq:rate_generic}
R_{u_l}^{(n)} = \frac{1}{2} B_{RB}  \log_2 \left( 1 + P_{u_l, l}^{(n)} \gamma_{u_l, l, 1}^{(n)} \right).
\end{equation}

\subsection{Continuous Relaxation and Reformulation}

The optimization problem $\mathbf{P1}$ is a mixed-integer non-linear program (MINLP) which is computationally intractable. A common approach to tackle this problem is to relax the constraint that an RB is used by only one UE by using the time-sharing factor \cite{relax-con-1}. Thus $x_{u_l}^{(n)} \in (0,1]$ is represented as the sharing factor where each $x_{u_l}^{(n)}$ denotes the portion of time that RB $n$ is assigned to UE $u_l$ and satisfies the constraint $\displaystyle \sum_{u_l \in \mathcal{U}_l} x_{u_l}^{(n)} \leq 1, ~\forall n$. Besides, we introduce a new variable $S_{u_l, l}^{(n)} = x_{u_l}^{(n)} P_{u_l,l}^{(n)}$ which denotes the actual transmit power of UE $u_l$ on RB $n$ \cite{time-share-1}. Then the relaxed problem can be stated as follows:  
\begin{subequations}
\setlength{\arraycolsep}{0.0em}
\begin{eqnarray}
 \mathbf{(P2)} \hspace{15em} \nonumber \\
\underset{x_{u_l}^{(n)}, S_{u_l, l}^{(n)}, \omega_{u_l}^{(n)}}{\operatorname{max}} ~ \sum_{u_l \in \mathcal{U}_l } \sum_{n =1}^N   \frac{1}{2}  x_{u_l}^{(n)}   B_{RB} \log_2  &&  \left(  1 +   \frac{S_{u_l, l}^{(n)} h_{u_l, l, 1}^{(n)}} {x_{u_l}^{(n)} \omega_{u_l}^{(n)}} \right) \hspace{0.1em}   \nonumber \\
\text{subject to} \quad \sum_{u_l \in \mathcal{U}_l} x_{u_l}^{(n)} ~ &&{\leq} ~ 1, \quad \forall n \hspace{-5em} \label{eq:con-bin-relx} \\
 \sum_{n =1}^N S_{u_l, l}^{(n)} ~ &&{\leq} ~ P_{u_l}^{max}, \forall u_l  \label{eq:con-pow-ue-relx} \\
 \sum_{u_l \in \mathcal{U}_l } \sum_{n =1}^N \frac{h_{u_l, l, 1}^{(n)}}{h_{l, u_l, 2}^{(n)}} S_{u_l, l}^{(n)} ~ &&{\leq} ~ P_l^{max} \label{eq:con-pow-rel-relx} \\
 \sum_{u_l \in \mathcal{U}_l } S_{u_l, l}^n g_{{u_l^*}, l, 1}^{(n)} ~ &&{\leq} ~ I_{th, 1}^{(n)}, ~~ \forall n \label{eq:con-intf-1-relx}\\
 \sum_{u_l \in \mathcal{U}_l } \frac{h_{u_l, l, 1}^{(n)}}{h_{l, u_l, 2}^{(n)}} S_{u_l, l}^{(n)} g_{l, {u_l^*}, 2}^{(n)} ~ &&{\leq} ~ I_{th, 2}^{(n)}, ~~\forall n \label{eq:con-intf-2-relx} \\
\quad \sum_{n=1}^N \frac{1}{2}  x_{u_l}^{(n)} B_{RB} \log_2 \left(  1 + \frac{S_{u_l, l}^{(n)} h_{u_l, l, 1}^{(n)}}{x_{u_l}^{(n)} \omega_{u_l}^{(n)}} \right)  ~ &&{\geq} ~ Q_{u_l},  ~~ \forall u_l   \label{eq:con-QoS-cue-relx}\\
\quad S_{u_l, l}^{(n)}  ~ &&{\geq} ~ 0,  ~~\forall n, u_l \label{eq:con-pow-0-relx} \\
 I_{u_l,l}^{(n)} + \sigma^2 ~ &&{\leq}~  \omega_{u_l}^{(n)}, \forall n, u_l \label{eq:con-aux-relx} 
\end{eqnarray}
\end{subequations}
where $\omega_{u_l}^{(n)}$ is an auxiliary variable for $u_l$ over RB $n$ and let $I_{u_l,l}^{(n)} = \max \left\lbrace  I_{u_l,l,1}^{(n)} , I_{l,u_l,2}^{(n)} \right\rbrace $. The duality gap of any optimization problem satisfying the time sharing condition is negligible as the number of RB becomes significantly large. Our optimization problem satisfies the time-sharing condition and hence the solution of the relaxed problem is asymptotically optimal \cite{large-rb-dual}. Since the objective function is concave, the constraint in (\ref{eq:con-QoS-cue-relx}) is convex, and all the remaining constraints are affine, the optimization problem $\mathbf{P2}$ is convex. Due to convexity of the optimization problem $\mathbf{P2}$, there exists a unique optimal solution. 

\begin{statement}
\label{theorem:power-rb-alloc-nominal}

(a) The power allocation for UE $u_l$ over RB $n$ is given by
\begin{equation}
\label{eq:power-alloc}
{P_{u_l, l}^{(n)}}^* = \frac{{S_{u_l,l}^{(n)}}^*}{{x_{u_l}^{(n)}}^*} = \left[\delta_{u_l,l}^{(n)} - \frac{\omega_{u_l}^{(n)}}{ h_{u_l,l, 1}^{(n)}}\right]^+
\end{equation} 
where $\delta_{u_l,l}^{(n)} = \frac{\tfrac{1}{2} B_{RB} \frac{(1 + \lambda_{u_l})}{\ln 2}}{\rho_{u_l} + \frac{h_{u_l,l, 1}^{(n)}}{h_{l, u_l,2}^{(n)}} \nu_l + g_{{u_l^*}, l, 1}^{(n)} \psi_{n} + \frac{h_{u_l, l, 1}^{(n)}}{h_{l, u_l, 2}^{(n)}} g_{l, {u_l^*}, 2}^{(n)} \varphi_{n} } $ and $[\epsilon]^+ = \max \left\lbrace \epsilon,0 \right\rbrace$.
\vspace{1em}

(b) The RB allocation is determined as follows:
\begin{equation}
\label{eq:channel_alloc1}
{x_{u_l}^{(n)}}^* = 
\begin{cases} 
\vspace{0.4em} 1, & \mu_n \leq \chi_{u_l,l}^{(n)} \\ 
 0, & \mu_n > \chi_{u_l,l}^{(n)} 
\end{cases}
\end{equation}
and $\chi_{u_l,l}^{(n)}$ is defined as 
\begin{equation}
 \chi_{u_l,l}^{(n)} = \tfrac{1}{2} (1+ \lambda_{u_l})  B_{RB} \left[  \log_2 \left(1 + \tfrac{S_{u_l,l}^{(n)} h_{u_l,l,1}^{(n)}}{x_{u_l}^{(n)} \omega_{u_l}^{(n)}} \right) - \theta_{u_l,l}^{(n)} \right]
\end{equation}
where $\theta_{u_l,l}^{(n)} = \tfrac{S_{u_l,l}^{(n)} \gamma_{u_l,l,1}^{(n)}}{\left(x_{u_l}^{(n)} \omega_{u_l}^{(n)} + S_{u_l,l}^{(n)} \gamma_{u_l,l,1}^{(n)} \right) \ln 2}$.

\end{statement}

\begin{IEEEproof}
See \textbf{Appendix \ref{app:power-rb-alloc-nominal}}.
\end{IEEEproof}

\begin{proposition}
\label{theorem:asyp_opt}
The power and RB allocation obtained by (\ref{eq:power-alloc}) and (\ref{eq:channel_alloc1}) is a globally optimal solution to the original problem $\mathbf{P1}$.
\end{proposition}

\begin{IEEEproof}
Since $\mathbf{P2}$ is a constraint-relaxed version of $\mathbf{P1}$, the solution $\left({x_{u_l}^{(n)}}^*, {P_{u_l, l}^{(n)}}^*\right)$
gives an upper bound to the objective of $\mathbf{P1}$. Besides, since ${x_{u_l}^{(n)}}^*$ satisfies the binary constraints in $\mathbf{P1}$, $\left({x_{u_l}^{(n)}}^*, {P_{u_l, l}^{(n)}}^*\right)$ satisfies all constraints in $\mathbf{P1}$ and hence also gives a lower bound.
\end{IEEEproof}

In the above problem formulation it is assumed that each of the relays and D2D UEs has the perfect information about the experienced interference. Also,  the channel gains between the relay and the other UEs (associated with neighbouring relays) are known to the relay. However, estimating  the exact values of link gains is not easy in practice. To deal with the uncertainties in the estimated values, we apply the worst-case robust optimization method \cite{robust-theroy}. 

\section{Robust Resource Allocation}
\label{sec:robust}

\subsection{Formulation of Robust Problem}

Let the vector of link gains between relay $l$ and other transmitting UEs (associated with other relays, i.e., for $ \forall j \in \mathcal{L}, j \neq l$) in the first hop over RB $n$ be denoted by $\mathbf{g}_{l, 1}^{(n)} = \left[g_{{1^*}, l, 1}^{(n)},~ g_{{2^*}, l, 1}^{(n)}, \cdots, ~ g_{{{|\mathcal{U}_l|}^*}, l, 1}^{(n)} \right]$, where $|\mathcal{U}_l|$ is the total number of UEs associated with relay $l$. Similarly, the vector of link gains between relay $l$ and receiving D2D UEs (associated with other relays) in the second hop over RB $n$ is given by $\mathbf{g}_{l, 2}^{(n)} = \left[g_{l, {1^*}, 2}^{(n)},~ g_{l, {2^*}, 2}^{(n)}, \cdots, ~ g_{{l, {|\mathcal{U}_l|}^*}, 2}^{(n)} \right]$.

We assume that the link gains and the aggregated interference (i.e., $I_{u_l, l}^{(n)}$, $\forall n, u_l$ and elements of $\mathbf{g}_{l, 1}^{(n)}, \mathbf{g}_{l, 2}^{(n)}$, $\forall n$) are unknown but  are bounded in a region (i.e., uncertainty set) with a given probability. 
For example, the channel gain in the first hop is bounded in $ \Re_{g_{l, 1}}^{(n)}$, with estimated value $\bar{\mathbf{g}}_{l, 1}^{(n)}$ and the bounded error $\hat{\mathbf{g}}_{l, 1}^{(n)}$, i.e., $\mathbf{g}_{l, 1}^{(n)} =  \bar{\mathbf{g}}_{l, 1}^{(n)} + \hat{\mathbf{g}}_{l, 1}^{(n)}$, and $\mathbf{g}_{l, 1}^{(n)} \in  \Re_{g_{l, 1}}^{(n)}, \forall n \in \mathcal{N}$, where $ \Re_{g_{l, 1}}^{(n)}$ is the uncertainty set for $\mathbf{g}_{l,1}^{(n)}$. Similarly, let $\Re_{g_{l, 2}}^{(n)},~ \forall n $ be the uncertainty set for the link gains in the second hop and $\Re_{I_{u_l,l}}^{(n)},~ \forall n, u_l $ be the uncertainty set for interference level. 

In the formulation of robust problem, we utilize a similar rate expression [i.e., equation (\ref{eq:rate_generic})] as the one used in the nominal problem formulation. Although dealing with similar utility function (i.e., rate equation) for both nominal and robust problems is quite common in literature (e.g., in \cite{uncertainity-cr-2, robust_rel, robust_intf}),  when perfect channel information is not available to receiver nodes, the rate obtained by (\ref{eq:rate_generic}) actually approximates the achievable rate \footnote{According to information-theoretic capacity analysis, in presence of channel uncertainties  at the receiver, the  lower and upper bounds of the rate are given by equations (46) and (49) in \cite{uncertainity-early}, respectively. However, for mathematical tractability, we resort to (\ref{eq:rate_generic}) to calculate the achievable data rate in both the nominal and robust problem formulations.} The solution to $\mathbf{P2}$ is robust against uncertainties if and only if for any realization of $\mathbf{g}_{l,1}^{(n)} \in \Re_{g_{l,1}}^{(n)}, \mathbf{g}_{l,2}^{(n)} \in \Re_{g_{l,2}}^{(n)} $, and $I_{u_l,l}^{(n)} \in \Re_{I_{u_l,l}}^{(n)}$, the optimal solution satisfies the constraints in (\ref{eq:con-intf-1-relx}),  (\ref{eq:con-intf-2-relx}), and (\ref{eq:con-aux-relx}). Therefore, the robust counterpart of $\mathbf{P2}$ is represented as  

\begin{subequations}
\setlength{\arraycolsep}{0.0em}
\begin{eqnarray}
 \mathbf{(P3)} \hspace{15em} \nonumber \\
 \underset{x_{u_l}^{(n)}, S_{u_l, l}^{(n)}, \omega_{u_l}^{(n)}}{\operatorname{max}} ~ \sum_{u_l \in \mathcal{U}_l } \sum_{n =1}^N   \frac{1}{2}  x_{u_l}^{(n)}   B_{RB} \log_2  &&  \left(  1 +   \frac{S_{u_l, l}^{(n)} h_{u_l, l, 1}^{(n)}}{x_{u_l}^{(n)} \omega_{u_l}^{(n)}} \right)     \nonumber \\
\text{subject to}~~	\text{(\ref{eq:con-bin-relx})},~
					\text{(\ref{eq:con-pow-ue-relx})},~  
					\text{(\ref{eq:con-pow-rel-relx})},~
					\text{(\ref{eq:con-intf-1-relx})},~ \nonumber \\
					\text{(\ref{eq:con-intf-2-relx})},&&~ 
  					\text{(\ref{eq:con-QoS-cue-relx})},~ 
					\text{(\ref{eq:con-pow-0-relx})},~
					\text{(\ref{eq:con-aux-relx})}  \nonumber \\
\text{and}~~ \mathbf{g}_{l,1}^{(n)} \in  \Re_{g_{l,1}}^{(n)}, ~~ \mathbf{g}_{l,2}^{(n)} ~ &&{\in} ~  \Re_{g_{l,2}}^{(n)}, ~~\forall n \label{eq:con-region-robst1} \\
 I_{u_l,l}^{(n)} ~ &&{\in} ~  \Re_{I_{u_l,l}}^{(n)}, ~\forall n, \forall u_l 
  \label{eq:con-region-intf-robst1}  \nonumber \\
\end{eqnarray}
\setlength{\arraycolsep}{5pt}
\end{subequations}
where the constraints in (\ref{eq:con-region-robst1}) and (\ref{eq:con-region-intf-robst1}) represent the requirements for the robustness of the solution. 
  
\begin{proposition}
\label{theorem:convex}
When $\Re_{g_{l,1}}^{(n)}, \Re_{g_{l,2}}^{(n)}$, and $\Re_{I_{u_l,l}}$ are compact and convex sets, $\mathbf{P3}$ is a convex optimization problem.
\end{proposition}

\begin{IEEEproof}
The uncertainty constraints in (\ref{eq:con-intf-1-relx}), (\ref{eq:con-intf-2-relx}), and (\ref{eq:con-aux-relx}) are satisfied if and only if 
\begin{align*}
\underset{\mathbf{g}_{l,1}^{(n)} \in  \Re_{g_{l,1}}^{(n)}}{\operatorname{max}} \sum_{u_l \in \mathcal{U}_l } S_{u_l, l}^{(n)} g_{{u_l^*}, l, 1}^{(n)} & \leq  I_{th, 1}^{(n)}, ~~ \forall n \\
\underset{\mathbf{g}_{l,2}^{(n)} \in  \Re_{g_{l,2}}^{(n)}}{\operatorname{max}} \sum_{u_l \in \mathcal{U}_l } \frac{h_{u_l, l, 1}^{(n)}}  {h_{l, u_l, 2}^{(n)}} S_{u_l, l}^{(n)} g_{l, {u_l^*}, 2}^{(n)}  & \leq  I_{th, 2}^{(n)}, ~~\forall n \\
\underset{I_{u_l,l}^{(n)} \in  \Re_{I_{u_l,l}}^{(n)}}{\operatorname{max}} I_{u_l,l}^{(n)} + \sigma^2 & \leq \omega_{u_l}^{(n)},  ~~\forall n, u_l
\end{align*}
which is equivalent to
\begin{align*}
 \sum_{u_l \in \mathcal{U}_l } S_{u_l, l}^{(n)} \bar{g}_{{u_l^*}, l, 1}^{(n)}  ~~ + \hspace{12em} \\ 
 \underset{\mathbf{g}_{l,1}^{(n)} \in  \Re_{g_{l,1}}^{(n)}}{\operatorname{max}} \sum_{u_l \in \mathcal{U}_l } S_{u_l, l}^{(n)} \left( g_{{u_l^*}, l, 1}^{(n)} -  \bar{g}_{{u_l^*}, l, 1}^{(n)} \right) \hspace*{1.7em} & \hspace*{-1.6em} \leq  I_{th, 1}^{(n)}, ~ \forall n \\
\sum_{u_l \in \mathcal{U}_l } \frac{h_{u_l, l, 1}^{(n)}}  {h_{l, u_l, 2}^{(n)}} S_{u_l, l}^{(n)} \bar{g}_{l, {u_l^*}, 2}^{(n)} ~~ + \hspace{9em} \\  
 \underset{\mathbf{g}_{l,2}^{(n)} \in  \Re_{g_{l,2}}^{(n)}}{\operatorname{max}} \sum_{u_l \in \mathcal{U}_l } \frac{h_{u_l, l, 1}^{(n)}}  {h_{l, u_l, 2}^{(n)}} S_{u_l, l}^{(n)} \left( g_{l, {u_l^*},  2}^{(n)} - \bar{g}_{l, {u_l^*}, 2}^{(n)} \right) \hspace*{1.7em}  & \hspace*{-1.6em} \leq  I_{th, 2}^{(n)}, ~\forall n \\
\bar{I}_{u_l,l}^{(n)} + \underset{I_{u_l,l}^{(n)} \in  \Re_{I_{u_l,l}^{(n)}}}{\operatorname{max}} \left( I_{u_l,l}^{(n)} - \bar{I}_{u_l,l}^{(n)} \right) + \sigma^2 \hspace*{1.7em} & \hspace*{-1.6em} \leq  \omega_{u_l}^{(n)}, ~\forall n, u_l.
\end{align*}

Since the $\max$ function over a convex set is a convex function (Section 3.2.4 in \cite{book-boyd}), convexity of the problem $\mathbf{P3}$ is conserved.
\end{IEEEproof}

The problem $\mathbf{P2}$ is the nominal problem of $\mathbf{P3}$ where it is assumed that the perfect channel state information is available, i.e., the estimated values are considered as exact values. With the inclusion of uncertainty in (\ref{eq:con-intf-1-relx}), (\ref{eq:con-intf-2-relx}), and (\ref{eq:con-aux-relx}), the constraints in the optimization problem $\mathbf{P3}$ are still affine. In order to express the constraints in closed-form (i.e., to avoid using the uncertainty set), in the following, we utilize the notion of protection function \cite{robust-theroy, general_norm} instead of uncertainty set.
 
\subsection{Uncertainty Set and Protection Function}

From $\mathbf{P3}$, the optimization problem is impacted by the uncertainty sets $\Re_{g_{l,1}}^{(n)}, \Re_{g_{l,2}}^{(n)}$, and $\Re_{I_{u_l,l}}^{(n)}$. To obtain the robust formulation, we consider that the uncertainty sets for the uncertain parameters are based on the differences between the actual (i.e., uncertain) and nominal (i.e., without considering uncertainty) values. These differences can be mathematically represented by general norms \cite{general_norm}. For example, the uncertainty sets for channel gain in the first and second hops for $\forall n \in \mathcal{N}$ are given by
\begin{subequations}
\setlength{\arraycolsep}{0.0em}
\begin{eqnarray}
 \Re_{g_{l,1}}^{(n)} ~ && {=}  ~ \left\lbrace \mathbf{g}_{l,1}^{(n)} | \parallel \mathbf{M}_{g_{l,1}}^{(n)} \cdot \left( \mathbf{g}_{l,1}^{(n)} - \bar{\mathbf{g}}_{l,1}^{(n)} \right)^\mathsf{T} \parallel \hspace{0.2em} \leq \Psi_{l,1}^{(n)} \right\rbrace \hspace{1.9em}  \label{eq:un_set_gain_1}\\
 \Re_{g_{l,2}}^{(n)} ~ && {=} ~ \left\lbrace \mathbf{g}_{l,2}^{(n)} | \parallel \mathbf{M}_{g_{l,2}}^{(n)} \cdot \left( \mathbf{g}_{l,2}^{(n)} - \bar{\mathbf{g}}_{l,2}^{(n)} \right)^\mathsf{T} \parallel \hspace{0.2em} \leq \Psi_{l,2}^{(n)} \right\rbrace \hspace{1.9em} \label{eq:un_set_gain_2}
\end{eqnarray}
\setlength{\arraycolsep}{5pt}
\end{subequations}
where  $\parallel \cdot \parallel$ denotes the general norm, $\Psi_{l,1}^{(n)}$ and $\Psi_{l,2}^{(n)}$ represent the bound on the uncertainty set; $ \mathbf{g}_{l,1}^{(n)}$, $\mathbf{g}_{l,2}^{(n)} $ are the actual and $ \bar{\mathbf{g}}_{l,1}^{(n)} $, $ \bar{\mathbf{g}}_{l,2}^{(n)}$ are the estimated (i.e., nominal) channel gain vectors; $\mathbf{M}_{g_{l,1}}^{(n)}$ and $\mathbf{M}_{g_{l,2}}^{(n)}$ are the invertible $\mathfrak{R}^{|\mathcal{U}_l| \times |\mathcal{U}_l|}$ weight matrices for the first and second hop, respectively. Likewise, the uncertainty set for the experienced interference is expressed as
\begin{equation}
\label{eq:uncertainity_reg_intf}
\Re_{I_{u_l,l}}^{(n)} = \left\lbrace I_{u_l,l}^{(n)} | \parallel M_{I_{u_l,l}}^{(n)} \cdot \left( I_{u_l,l}^{(n)} - \bar{I}_{u_l,l}^{(n)} \right) \parallel \hspace{0.2em} \leq \Upsilon_{u_l}^{(n)} \right\rbrace
\end{equation}    
where $ I_{u_l,l}^{(n)}$ and $\bar{I}_{u_l,l}^{(n)} $ are the actual  and estimated interference levels, respectively; the variable $M_{I_{u_l,l}}^{(n)}$ denotes weight and  $\Upsilon_{u_l}^{(n)}$ is the upper bound on the uncertainty set.

In the proof of \textbf{Proposition \ref{theorem:convex}}, the terms 
\begin{subequations}
\setlength{\arraycolsep}{0.0em}
\begin{eqnarray}
  \Delta_{g_{l, 1}}^{(n)} ~ &&{=} ~ \underset{\mathbf{g}_{l,1}^{(n)} \in  \Re_{g_{l,1}}^{(n)} }{\operatorname{max}} \displaystyle \sum_{u_l \in \mathcal{U}_l } S_{u_l, l}^{(n)} \left( g_{{u_l^*}, l, 1}^{(n)} -  \bar{g}_{{u_l^*}, l, 1}^{(n)} \right) \label{eq:prot-func-gain-1} \\
  \Delta_{g_{l, 2}}^{(n)}  ~ &&{=} ~ \underset{\mathbf{g}_{l,2}^{(n)} \in  \Re_{g_{l,2}}^{(n)} }{\operatorname{max}} \displaystyle \sum_{u_l \in \mathcal{U}_l } \tfrac{h_{u_l, l, 1}^{(n)}}  {h_{l, u_l, 2}^{(n)}} S_{u_l, l}^{(n)} \left( g_{l, {u_l^*}, 2}^{(n)} - \bar{g}_{l, {u_l^*}, 2}^{(n)} \right) \label{eq:prot-func-gain-2} 
\\
 \Delta_{I_{u_l,l}}^{(n)} ~ &&{=} ~ \underset{I_{u_l,l}^{(n)} \in  \Re_{I_{u_l,l}}^{(n)}}{\operatorname{max}} \left( I_{u_l,l}^{(n)} - \bar{I}_{u_l,l}^{(n)} \right) \label{eq:prot-func-intf}
\end{eqnarray}
\setlength{\arraycolsep}{5pt}
\end{subequations} 
are called protection functions for constraint (\ref{eq:con-intf-1-relx}), (\ref{eq:con-intf-2-relx}), and (\ref{eq:con-aux-relx}), respectively, whose value (i.e., protection value) depends on the uncertain parameters. Using the protection function, the optimization problem can be rewritten as
\begin{subequations}
\setlength{\arraycolsep}{0.0em}
\begin{eqnarray}
 \mathbf{(P4)} \hspace{15em} \nonumber \\ 
\underset{x_{u_l}^{(n)}, S_{u_l, l}^{(n)}, \omega_{u_l}^{(n)}}{\operatorname{max}} ~ \sum_{u_l \in \mathcal{U}_l } \sum_{n =1}^N   \frac{1}{2}  x_{u_l}^{(n)}   B_{RB}  \log_2  && \left(  1 +   \frac{S_{u_l, l}^{(n)} h_{u_l, l, 1}^{(n)}}{x_{u_l}^{(n)} \omega_{u_l}^{(n)}} \right)    \nonumber \\
\text{subject to}~~	\text{(\ref{eq:con-bin-relx})},~
					\text{(\ref{eq:con-pow-ue-relx})},~  
					\text{(\ref{eq:con-pow-rel-relx})},&&~ 
  					\text{(\ref{eq:con-QoS-cue-relx})},~ 
					\text{(\ref{eq:con-pow-0-relx})}~~\text{and} \nonumber \\
\sum_{u_l \in \mathcal{U}_l } S_{u_l, l}^{(n)} \bar{g}_{{u_l^*}, l, 1}^{(n)} + \Delta_{g_{l, 1}}^{(n)} ~ &&{\leq} ~ I_{th, 1}^{(n)}, ~~ \forall n \label{eq:con-intf-1-robst2}\\
\sum_{u_l \in \mathcal{U}_l } \frac{h_{u_l, l, 1}^{(n)}}  {h_{l, u_l, 2}^{(n)}} S_{u_l, l}^{(n)} \bar{g}_{l, {u_l^*}, 2}^{(n)} + \Delta_{g_{l, 2}}^{(n)} ~ &&{\leq} ~ I_{th, 2}^{(n)}, ~~\forall n \label{eq:con-intf-2-robst2} \\
\bar{I}_{u_l,l}^{(n)} + \Delta_{I_{u_l,l}}^{(n)} + \sigma^2 ~ &&{\leq} ~ \omega_{u_l}^{(n)},  \forall n, u_l \label{eq:con-aux-robst2} 
\end{eqnarray}
\setlength{\arraycolsep}{5pt}
\end{subequations}
where $\Delta_{g_{l, 1}}^{(n)}, \Delta_{g_{l, 2}}^{(n)}$, and $\Delta_{I_{u_l,l}}^{(n)} $ are defined by (\ref{eq:prot-func-gain-1}), (\ref{eq:prot-func-gain-2}), and (\ref{eq:prot-func-intf}), respectively.

\begin{proposition}
\label{theorem:norm}
The protection functions for the uncertainty sets represented by general norms [i.e., by (\ref{eq:un_set_gain_1}), (\ref{eq:un_set_gain_2}), and (\ref{eq:uncertainity_reg_intf})] are
\begin{subequations}
\setlength{\arraycolsep}{0.0em}
\begin{align}
\Delta_{g_{l, 1}}^{(n)} &= \Psi_{l,1}^{(n)}  \parallel {\mathbf{M}_{g_{l,1}}^{(n)}}^{-1} \cdot \left(\mathbf{S}_{l,1}^{(n)} \right)^\mathsf{T} \parallel^* \\
\Delta_{g_{l, 2}}^{(n)} &= \Psi_{l,2}^{(n)} \parallel {\mathbf{M}_{g_{l,2}}^{(n)}}^{-1} \cdot \left(\mathbf{H}_{l}^{(n)}  \cdot \mathbf{S}_{l,1}^{(n)}  \right)^\mathsf{T} \parallel^*  \\
\Delta_{I_{u_l,l}}^{(n)} &= \Upsilon_{u_l}^{(n)} \parallel {M_{I_{u_l,l}}^{(n)}}^{-1} \cdot I_{u_l,l}^{(n)} \parallel^* 
\end{align}
\setlength{\arraycolsep}{5pt}
\end{subequations}
where $\mathbf{S}_{l, 1}^{(n)} = \left[S_{{1}, l}^{(n)},~ S_{{2}, l}^{(n)}, \cdots, ~ S_{{{|\mathcal{U}_l|}}, l}^{(n)} \right]$, $\mathbf{H}_{l}^{(n)} = \left[ \tfrac{h_{1, l, 1}^{(n)}}  {h_{l, 1, 2}^{(n)}}, ~ \tfrac{h_{2, l, 1}^{(n)}}  {h_{l, 2, 2}^{(n)}} , \cdots, ~ \tfrac{h_{|\mathcal{U}_l|, l, 1}^{(n)}}  {h_{l, |\mathcal{U}_l|, 2}^{(n)}} \right] $ and $\parallel \cdot \parallel^*$ is the dual norm of $\parallel \cdot \parallel$.

\end{proposition} 

\begin{IEEEproof}
Using the expression $\mathbf{w}_{l,1}^{(n)} = \frac{\mathbf{M}_{g_{l,1}}^{(n)} \cdot \left( \bar{\mathbf{g}}_{l,1}^{(n)} - \mathbf{g}_{l,1}^{(n)} \right)^\mathsf{T}}{\Psi_{l,1}^{(n)}}$, the uncertainty set (\ref{eq:un_set_gain_1}) becomes
\begin{equation}
\Re_{g_l,1}^{(n)} = \left\lbrace \mathbf{w}_{l,1}^{(n)} \vert \parallel \bar{\mathbf{w}}_{l,1}^{(n)} - \mathbf{w}_{l,1}^{(n)} \parallel \hspace{0.2em} \leq 1 \right\rbrace, \quad \forall n.
\end{equation}
Besides, the protection function (\ref{eq:prot-func-gain-1}) can be rewritten as 
\begin{eqnarray}
\underset{\mathbf{g}_{l,1}^{(n)} \in  \Re_{g_{l,1}}^{(n)} }{\operatorname{max}} \hspace{-1.5em} & \displaystyle \sum_{u_l \in \mathcal{U}_l } S_{u_l, l}^{(n)}  \left( g_{{u_l^*}, l, 1}^{(n)} -  \bar{g}_{{u_l^*}, l, 1}^{(n)} \right) \nonumber \\ =& \underset{\mathbf{g}_{l,1}^{(n)} \in  \Re_{g_{l,1}}^{(n)} }{\operatorname{max}}  \mathbf{S}_{l, 1}^{(n)} \cdot \left( \mathbf{g}_{l,1}^{(n)} -  \bar{\mathbf{g}}_{l,1}^{(n)}\right)^\mathsf{T} \nonumber \\ =& \underset{\mathbf{g}_{l,1}^{(n)} \in  \Re_{g_{l,1}}^{(n)} }{\operatorname{max}}  \mathbf{S}_{l, 1}^{(n)} \cdot \left( {\mathbf{M}_{g_{l,1}}^{(n)}}^{-1} \cdot \mathbf{w}_{l,1}^{(n)} \right).
\label{eq:prtection_func_proof}
\end{eqnarray}

Note that, given a norm $\parallel \mathbf{y} \parallel$ for a vector $\mathbf{y}$, its dual norm induced over the dual space of linear
functionals $\mathbf{z}$ is $\parallel \mathbf{z} \parallel^* = \underset{\parallel \mathbf{y} \parallel \leq 1}{\operatorname{max}} ~\mathbf{z}^\mathsf{T} \mathbf{y}$ \cite{general_norm}. Since the protection function in (\ref{eq:prtection_func_proof}) is the dual norm of uncertainty region in (\ref{eq:un_set_gain_1}), the proof follows. The protection functions for the uncertainity sets in (\ref{eq:un_set_gain_2}) and (\ref{eq:uncertainity_reg_intf}) are obtained in a similar way. 
\end{IEEEproof}


Since the dual norm is a convex function, the convexity of $\mathbf{P4}$ is preserved. In addition, when the uncertainty set for any vector $\mathbf{y}$ is a linear norm defined by ${\parallel \mathbf{y} \parallel}_\alpha = \left( \sum \mathbf{abs}\lbrace y \rbrace^\alpha \right)^{\frac{1}{\alpha}}$ with order $\alpha \geq 2$, where $\mathbf{abs}\lbrace y \rbrace$ is the absolute value of $y$ and the dual norm is a linear norm with order $\beta = 1+ \frac{1}{\alpha-1}$. In such cases, the protection function can be defined as a linear norm of order $\beta$. Therefore, the protection function becomes a deterministic function of the optimization variables (i.e., $x_{u_l}^{(n)}, S_{u_l, l}^{(n)}$, and $\omega_{u_l}^{(n)}$), and the non-linear $\max$ function is eliminated from the protection functions [i.e., from  constraint (\ref{eq:con-intf-1-robst2}), (\ref{eq:con-intf-2-robst2}), and (\ref{eq:con-aux-robst2})]. Consequently, the resource allocation problem turns out to be a standard form of convex optimization problem $\mathbf{P5}$, where $\Delta_{I_{u_l,l}}^{(n)} = \Upsilon_{u_l}^{(n)} {\parallel {M_{I_{u_l,l}}^{(n)}}^{-1} \cdot I_{u_l,l}^{(n)} \parallel}_\beta $ and $\mathbf{A}(j,:)$ denotes the $j$-th row of matrix $\mathbf{A}$.

In the LTE-A system, which exploits orthogonal frequency-division multiplexing (OFDM) for radio access, fading can be considered uncorrelated across RBs (Chapter 1 in \cite{book:fading_uncor}); hence, it can be assumed that uncertainty and channel gain in each element of $\mathbf{g}_{l,1}^{(n)}$ and $\mathbf{g}_{l,2}^{(n)}$  are i.i.d. random variables \cite{fading_uncor2}. Therefore, $\mathbf{M}_{g_{l,1}}^{(n)}$ and $\mathbf{M}_{g_{l,2}}^{(n)}$ become a diagonal matrix. Note that for any diagonal matrix $\mathbf{A}$ with $j$-th diagonal element $a_{jj}$, the vector ${\mathbf{A}}^{-1}(j,:)$ contains only non-zero elements $\frac{1}{a_{jj}} $. In addition, since the channel uncertainties  are random, a commonly used approach is to represent the uncertainty set by an ellipsoid, i.e., the linear norm with $\alpha = 2$ so that the dual norm is a linear norm with $\beta =2$ \cite{ellip-m2, ellip-m2-1}. Hence, problem $\mathbf{P5}$ turns to a conic quadratic programming problem \cite{book:conic}. In order to solve $\mathbf{P5}$ efficiently, a distributed gradient-aided algorithm is developed in the following section.

\begin{figure*}[!t]
\normalsize

\begin{subequations}
\setlength{\arraycolsep}{0.0em}
\begin{eqnarray}
 \mathbf{(P5)} \hspace{2em}  
\underset{x_{u_l}^{(n)}, S_{u_l, l}^{(n)}, \omega_{u_l}^{(n)}}{\operatorname{max}} ~ \sum_{u_l \in \mathcal{U}_l } \sum_{n =1}^N   \frac{1}{2}  x_{u_l}^{(n)}   B_{RB}  \log_2  && \left(  1 +   \frac{S_{u_l, l}^{(n)} h_{u_l, l, 1}^{(n)}}{x_{u_l}^{(n)} \omega_{u_l}^{(n)}} \right)    \nonumber \\
\text{subject to} \quad \sum_{u_l \in \mathcal{U}_l} x_{u_l}^{(n)} ~ &&{\leq} ~ 1, \quad \quad ~ \forall n \hspace{-5em} \label{eq:con-bin-robst3} \\
\sum_{n =1}^N S_{u_l, l}^{(n)} ~ &&{\leq} ~ P_{u_l}^{max}, ~\forall u_l  \label{eq:con-pow-ue-robst3} \\
\sum_{u_l \in \mathcal{U}_l } \sum_{n =1}^N \frac{h_{u_l, l, 1}^{(n)}}{h_{l, u_l, 2}^{(n)}} S_{u_l, l}^{(n)} ~ &&{\leq} ~ P_l^{max} \label{eq:con-pow-rel-robst3} \\
\sum_{u_l \in \mathcal{U}_l } S_{u_l, l}^{(n)} \bar{g}_{{u_l^*}, l, 1}^{(n)} +  \Psi_{l,1}^{(n)} \left( \sum_{k=1}^{|\mathcal{U}_l|} \left( {\mathbf{M}_{g_{l,1}}^{(n)}}^{-1}(k,:) \cdot \mathbf{S}_{l, 1}^{(n)} \right)^\beta  \right)^{\frac{1}{\beta}}  ~ &&{\leq} ~ I_{th, 1}^{(n)}, ~~ \forall n \label{eq:con-intf-1-robst3}\\
\sum_{u_l \in \mathcal{U}_l } \frac{h_{u_l, l, 1}^{(n)}}  {h_{l, u_l, 2}^{(n)}} S_{u_l, l}^{(n)} \bar{g}_{l, {u_l^*}, 2}^{(n)} + \Psi_{l,2}^{(n)} \left( \sum_{k=1}^{|\mathcal{U}_l|} \left( {\mathbf{M}_{g_{l,2}}^{(n)}}^{-1}(k,:) \cdot \left( \mathbf{H}_{l}^{(n)} \cdot \mathbf{S}_{l, 1}^{(n)} \right) \right)^\beta  \right)^{\frac{1}{\beta}} ~ &&{\leq} ~ I_{th, 2}^{(n)}, ~~\forall n \label{eq:con-intf-2-robst3} \\
\sum_{n=1}^N \frac{1}{2}  x_{u_l}^{(n)} B_{RB} \log_2 \left(  1 + \frac{S_{u_l, l}^{(n)} h_{u_l, l, 1}^{(n)}}{x_{u_l}^{(n)} \omega_{u_l}^{(n)}} \right)  ~ &&{\geq} ~ Q_{u_l},  ~~~ \forall u_l   \label{eq:con-QoS-cue-robst3}\\
S_{u_l, l}^{(n)}  ~ &&{\geq} ~ 0,  ~~~~~~\forall n, u_l\label{eq:con-pow-0-robst3} \\
\bar{I}_{u_l,l}^{(n)} + \Delta_{I_{u_l,l}}^{(n)} + \sigma^2 ~ &&{\leq}~  \omega_{u_l}^{(n)}, ~~ \forall n, u_l \label{eq:con-aux-robst3} 
\end{eqnarray}
\setlength{\arraycolsep}{5pt}
\end{subequations}

\hrulefill
\vspace*{4pt}
\end{figure*}

\section{Robust Distributed Algorithm} \label{sec:robust_algo}

\subsection{Algorithm Development}

\begin{statement}
\label{theorem:robust-alloc}
(a) The optimal power allocation for $u_l$ over RB $n$ is given by the following water-filling equation:
\begin{equation}
\label{eq:power-alloc-robust}
{P_{u_l, l}^{(n)}}^* = \frac{{S_{u_l,l}^{(n)}}^*}{{x_{u_l}^{(n)}}^*} = \left[\delta_{u_l,l}^{(n)} - \frac{\omega_{u_l}^{(n)}}{ h_{u_l,l, 1}^{(n)}}\right]^+
\end{equation} 
where $\delta_{u_l,l}^{(n)}$ is found by (\ref{eq:robust-del-pow}).

\begin{figure*}[!t]
\normalsize

\begin{equation}
\label{eq:robust-del-pow}
\delta_{u_l,l}^{(n)} = \frac{\tfrac{1}{2} B_{RB} \frac{(1 + \lambda_{u_l})}{\ln 2}}{\rho_{u_l} + \nu_l \frac{h_{u_l,l, 1}^{(n)}}{ h_{l, u_l,2}^{(n)}}  + \psi_{n} \left( \bar{g}_{{u_l^*}, l, 1}^{(n)} + \Psi_{l,1}^{(n)} m_{{u_l}{u_l}_{g_{l,1}}}^{(n)} \right)  + \varphi_{n} \frac{h_{u_l, l, 1}^{(n)}}{h_{l, u_l, 2}^{(n)}} \left(  \bar{g}_{l, {u_l^*}, 2}^{(n)} + \Psi_{l,2}^{(n)} m_{{u_l}{u_l}_{g_{l,2}}}^{(n)} \right) }
\end{equation}

\hrulefill
\vspace*{4pt}
\end{figure*}

(b) The RB allocation for $u_l$ over RB $n$ is obtained by (\ref{eq:channel_alloc1}).
\end{statement} 

\begin{IEEEproof}
See\textbf{ Appendix \ref{app:power-rb-alloc-robust}}. 
\end{IEEEproof}

Based on \textbf{Statement \ref{theorem:robust-alloc}}, we utilize a gradient-based method (given in \textbf{Appendix \ref{app:lagrange_update}}) to update the variables. Each relay independently performs the resource allocation and allocates resources to the associated UEs. For completeness, the distributed joint RB and power allocation algorithm is summarized in \textbf{Algorithm \ref{alg:rec_alloc}}.  

\begin{algorithm}
\caption{Joint RB and power allocation algorithm}
\label{alg:rec_alloc}
\begin{algorithmic}[1]   

\STATE  Each relay  $l \in \mathcal{L}$ estimates the reference gain $\bar{g}_{u_l^*, l, 1}^{(n)}$ and  $\bar{g}_{u_l^*, l, 2}^{(n)}$ from previous time slot $\forall u_l \in \mathcal{U}_l$ and $n \in \mathcal{N}$.

\STATE Initialize Lagrange multipliers to some positive value and set $t:=0$, $S_{u_l,l}^{(n)} := \frac{P_{u_l}^{max}}{N}$ $\forall u_l, n$.

\REPEAT 

\STATE Set $t:= t + 1$.

\STATE Calculate $x_{u_l}^{(n)}$ and $S_{u_l, l}^{(n)}$ for $\forall u_l, n$ using  (\ref{eq:channel_alloc1}) and (\ref{eq:power-alloc-robust}).

\STATE Update the Lagrange multipliers by (\ref{eq:lagrange_update1})--(\ref{eq:lagrange_update8}) and calculate the aggregated achievable network rate as $\displaystyle R_l(t) := \sum_{u_l \in \mathcal{U}_l} R_{u_l}(t)$.

\UNTIL $t = T_{max}$ or the convergence criterion met (i.e., $ \mathbf{abs} \lbrace R_l(t) - R_l(t-1) \rbrace < \varepsilon$, where $\varepsilon$ is the  tolerance for convergence).

\STATE  Allocate resources (i.e., RB and transmit power) to associated UEs for each relay and calculate the average achievable data rate.

\end{algorithmic}
\end{algorithm}

Note that, the L3 relays are able to perform their own scheduling (unlike L1 and L2 relays in \cite{relay-book-1}) as an eNB. These relays can obtain information such as the transmission power allocation at the other relays, channel gain information, etc. by using the X2 interface (Section 7 in \cite{lte_arch}) defined in the 3GPP specifications. In particular, a separate load indication procedure is used over the X2 interface for interface management (for details refer to \cite{lte_arch} and references therein). As a result, the relays can obtain the channel state information without increasing signaling overhead at the eNB. 

\subsection{Complexity Analysis}

\begin{proposition}
Using a small step size in gradient-based updating, the proposed algorithm achieves a sum-rate such that the difference in the sum rate in successive iterations is less than an arbitrary $\varepsilon >0$ with a polynomial computation complexity in  $|\mathcal{U}_l|$ and $N$.
\end{proposition}

\begin{IEEEproof}
It is easy to verify that the computational complexity at each iteration of  variable updating in (\ref{eq:lagrange_update1})--(\ref{eq:lagrange_update8}) is polynomial in  $|\mathcal{U}_l|$ and $N$. There are $|\mathcal{U}_l|N$ computations which are required to obtain the reference gains and if $T$ iterations are required for convergence, the overall complexity of the algorithm is $\mathcal{O}\left(|\mathcal{U}_l| N + T|\mathcal{U}_l| N\right)$. 

For any Lagrange multiplier $\kappa$, if we choose $\kappa(0)$ in the interval $[0, \kappa_{max}]$, the distance between $\kappa(0)$ and $\kappa^*$ is upper bounded by $\kappa_{max}$. Then it can be shown that at iteration $t$, the distance between the current best objective and the optimum objective is upper bounded by $\frac{\kappa_{max}^2 + \kappa(t)^2 \displaystyle \sum_{i=i}^{t} {\Lambda_{\kappa}^{(i)}}^2}{2 \displaystyle \sum_{i=i}^{t}\Lambda_{\kappa}^{(i)}}.$ If we take the step size $\Lambda_{\kappa}^{(i)} = \frac{a}{\sqrt{i}}$, where $a$ is a small constant, there are $\mathcal{O}\left(\frac{1}{\varepsilon^2}\right)$ iterations required for convergence to have the bound less than $\varepsilon$ \cite{notes_subgrad}. Hence, the complexity of the proposed algorithm is $\mathcal{O}\left( \left(1+ \tfrac{1}{\varepsilon^2} \right) |\mathcal{U}_l| N\right)$.
\end{IEEEproof}


\subsection{Cost of Robust Resource Allocation}

An important issue in robust resource allocation is the substantial reduction in the achievable network sum-rate. Reduction of achievable sum-rate due to introducing robustness is measured by $\mathscr{R}_\Delta = {\parallel R^* - R_\Delta^* \parallel}_2$, where $R^*$ and $R_\Delta^*$ are the optimal achievable sum-rates obtained by solving the nominal and the robust problem, respectively.

\begin{proposition}
\label{theorem:robust-tradeoff}

Let $\boldsymbol{\psi^*}$, $\boldsymbol{\varphi^*}$, $\boldsymbol{\varrho^*}$ be the optimal values of Lagrange multipliers for constraint (\ref{eq:con-intf-1-relx}), (\ref{eq:con-intf-2-relx}), and (\ref{eq:con-aux-relx}) in $\mathbf{P2}$, respectively. For all values of $\Delta_{g_{l, 1}}^{(n)}, \Delta_{g_{l, 2}}^{(n)}$, and $\Delta_{I_{u_l, l}}^{(n)}$ the reduction of achievable sum rate can be approximated as 
\begin{equation}
\label{eq:robust-opt-tradeoff}
\mathscr{R}_\Delta \approx \sum_{n=1}^{N} \psi_n^* \Delta_{g_{l, 1}}^{(n)} +  \sum_{n=1}^{N} \varphi_n^* \Delta_{g_{l, 2}}^{(n)} +  \sum_{u_l \in \mathcal{U}_l}\sum_{n=1}^{N} \varrho_{u_l}^{n*} \Delta_{I_{u_l, l}}^{(n)}.
\end{equation}

\end{proposition}

\begin{IEEEproof}
See \textbf{Appendix \ref{app:sensitivity}}.
\end{IEEEproof}

From \textbf{Proposition \ref{theorem:robust-tradeoff}}, the value of $\mathscr{R}_\Delta$ depends on the uncertainty set and by adjusting the size of $\Delta_{g_{l, 1}}^{(n)}$ and $\Delta_{g_{l, 2}}^{(n)}$,  $\mathscr{R}_\Delta$ can be controlled.

\subsection{Trade-off Between Robustness and Achievable Sum-rate}

The robust worst-case resource allocation dealing with channel uncertainties is very conservative and often leads to inefficient utilization of resources. In practice, uncertainty does not always correspond to its worst-case and in many instances the robust worst-case resource allocation may not be necessary. In such cases, it is desirable to achieve a trade-off between robustness and network sum-rate. This can be achieved through modifying the worst-case approach, where the uncertainty set is chosen in such a way that the probability of violating the interference threshold in both the hops is kept below a predefined level, and the network sum-rate is kept close to optimal value of nominal case. Therefore, we modify the constraints (\ref{eq:con-intf-1-relx}) and  (\ref{eq:con-intf-2-relx}) in $\mathbf{P2}$ as
\begin{subequations}
\setlength{\arraycolsep}{0.0em}
\begin{eqnarray}
 \mathbb{P} \left( \sum_{u_l \in \mathcal{U}_l } S_{u_l, l}^{(n)} g_{{u_l^*}, l, 1}^{(n)}  \geq  I_{th, 1}^{(n)} \right) \leq \Theta_{l,1}^{(n)}, ~~ \forall n \hspace{2em} \label{eq:chance-inft-1}\\
\mathbb{P} \left( \sum_{u_l \in \mathcal{U}_l } \frac{h_{u_l, l, 1}^{(n)}}{h_{l, u_l, 2}^{(n)}} S_{u_l, l}^{(n)} g_{l, {u_l^*}, 2}^{(n)} \geq I_{th, 2}^{(n)} \right) \leq \Theta_{l,2}^{(n)}, ~~\forall n \hspace{2em} \label{eq:chance-inft-2}
\end{eqnarray}
\end{subequations}
where $\Theta_{l,1}^{(n)}$ and $\Theta_{l,2}^{(n)}$ are given probabilities of violation of constraints (\ref{eq:con-intf-1-relx})  and (\ref{eq:con-intf-2-relx}) for any $n$ in the first hop and second hop, respectively. By changing $\Theta_{l,1}^{(n)}$ and $\Theta_{l,2}^{(n)}$, the trade-off between robustness and optimality will be achieved. By reducing $\Theta_{l,1}^{(n)}$ and $\Theta_{l,2}^{(n)}$, the network becomes more robust against uncertainty, while by increasing $\Theta_{l,1}^{(n)}$ and $\Theta_{l,2}^{(n)}$, the network sum-rate is increased. 

To deal with this trade-off we use the \textit{chance constrained approach}. When the constraints are affine functions, for i.i.d. values of uncertain parameters, (\ref{eq:con-intf-1-relx}) and (\ref{eq:con-intf-2-relx}) can be replaced by convex functions as their safe approximations \cite{robust-theroy}. Applying this approach we obtain 
\begin{subequations}
\setlength{\arraycolsep}{0.0em}
\begin{eqnarray*}
\hspace{-1em} \sum_{u_l \in \mathcal{U}_l } S_{u_l, l}^{(n)} g_{{u_l^*}, l, 1}^{(n)} =  \sum_{u_l \in \mathcal{U}_l } S_{u_l, l}^{(n)} \bar{g}_{{u_l^*}, l, 1}^{(n)} + \sum_{u_l \in \mathcal{U}_l } \xi_{u_l, l,1}^{(n)} S_{u_l, l}^{(n)} \hat{g}_{{u_l^*}, l, 1}^{(n)}   \label{eq:chance-aprx-inft-1}\\
 \sum_{u_l \in \mathcal{U}_l } \tfrac{h_{u_l, l, 1}^{(n)}}{h_{l, u_l, 2}^{(n)}} S_{u_l, l}^{(n)} g_{l,{u_l^*}, 2}^{(n)} = \hspace{14.5em} \\
 \sum_{u_l \in \mathcal{U}_l } \tfrac{h_{u_l, l, 1}^{(n)}}{h_{l, u_l, 2}^{(n)}} S_{u_l, l}^{(n)} \bar{g}_{l, {u_l^*}, 2}^{(n)} + \sum_{u_l \in \mathcal{U}_l } \xi_{l, u_l,2}^{(n)} \tfrac{h_{u_l, l, 1}^{(n)}}{h_{l, u_l, 2}^{(n)}} S_{u_l, l}^{(n)} \hat{g}_{l, {u_l^*},  2}^{(n)} \hspace{1em} \label{eq:chance-aprx-inft-2}
\end{eqnarray*}
\end{subequations}
where $\xi_{j}^{(n)} = \frac{g_{j}^{(n)} - \bar{g}_{j}^{(n)}}{\hat{g}_{j}^{(n)}}, \forall n$  is varied within the range $[-1,+1]$.
Under the assumption of uncorrelated fading channels, all values of $\xi_{u_l,l,1}^{(n)}$ and $\xi_{l, u_l,2}^{(n)}$ are independent of each other and belong to a specific class of probability distribution $\mathcal{P}_{u_l,l,1}^{(n)}$ and $\mathcal{P}_{l, u_l,2}^{(n)}$, respectively. Now the constraints in (\ref{eq:con-intf-1-relx}) and (\ref{eq:con-intf-2-relx}) can be replaced by Bernstein approximations of chance constraints  \cite{robust-theroy} as follows:
\begin{subequations}
\setlength{\arraycolsep}{0.0em}
\begin{eqnarray}
 \sum_{u_l \in \mathcal{U}_l } S_{u_l, l}^{(n)} \bar{g}_{{u_l^*}, l, 1}^{(n)} + \tilde{\Delta}_{g_{l, 1}}^{(n)} \leq  I_{th, 1}^{(n)} , ~~ \forall n \hspace{2em} \label{eq:berns-cons-inft-1}\\
  \sum_{u_l \in \mathcal{U}_l } \tfrac{h_{u_l, l, 1}^{(n)}}{h_{l, u_l, 2}^{(n)}} S_{u_l, l}^{(n)} \bar{g}_{l, {u_l^*}, 2}^{(n)} + \tilde{\Delta}_{g_{l, 2}}^{(n)} \leq  I_{th, 2}^{(n)} , ~~ \forall n \hspace{2em} \label{eq:berns-cons-inft-2}
\end{eqnarray}
\end{subequations}
where the protection functions $\tilde{\Delta}_{g_{l, 1}}^{(n)}$ and $\tilde{\Delta}_{g_{l, 2}}^{(n)}$ are given by (\ref{eq:berns-inft-1}) and (\ref{eq:berns-inft-1}), respectively. The variables $-1 \leq \eta_{\mathcal{P}_{j}}^{+} \leq +1$ and $\tau_{\mathcal{P}_{j}} \geq 0$ are used for safe approximation of chance constraints and depend on the probability distribution $\mathcal{P}_{j}$. For a fixed value of $\mathcal{P}_{j}$ the values of these parameters are listed in Table \ref{tab:berns_val} (see \textbf{Appendix \ref{app:berns_val}}). The constraints in (\ref{eq:berns-cons-inft-1}) and (\ref{eq:berns-cons-inft-2}) turn the resource allocation problem into a conic quadratic programming problem \cite{book:conic} and using the inequality ${\parallel \mathbf{y} \parallel}_2 \leq {\parallel \mathbf{y} \parallel}_1$, the  optimal RB and power allocation can be obtained in a distributed manner similar to that in \textbf{Algorithm \ref{alg:rec_alloc}}. Note that in (\ref{eq:berns-inft-1}) and (\ref{eq:berns-inft-2}), the protection functions depend on $\Theta_{l,1}^{(n)}$ and $\Theta_{l,2}^{(n)}$. By adjusting $\Theta_{l,1}^{(n)}$ and $\Theta_{l,2}^{(n)}$, a trade-off between rate and robustness can be achieved.

\begin{figure*}[!t]
\normalsize

\begin{subequations}
\setlength{\arraycolsep}{0.0em}
\begin{eqnarray}
 \tilde{\Delta}_{g_{l, 1}}^{(n)} =  \sum_{u_l \in \mathcal{U}_l } \eta_{\mathcal{P}_{u_l,l,1}^{(n)}}^{+} S_{u_l, l}^{(n)} \hat{g}_{{u_l^*}, l, 1}^{(n)} 
 + \sqrt{2 \ln \tfrac{1}{\Theta_{l,1}^{(n)}}} \left( \sum_{u_l \in \mathcal{U}_l } \tau_{\mathcal{P}_{u_l,l,1}^{(n)}}^2 \left( S_{u_l, l}^{(n)} \hat{g}_{{u_l^*}, l, 1}^{(n)} \right)^2 \right)^{\tfrac{1}{2}} , ~~ \forall n \hspace{2em} \label{eq:berns-inft-1}\\
  \tilde{\Delta}_{g_{l, 2}}^{(n)} = \sum_{u_l \in \mathcal{U}_l } \eta_{\mathcal{P}_{l, u_l,2}^{(n)}}^{+} S_{u_l, l}^{(n)} \hat{g}_{l, {u_l^*}, 2}^{(n)} 
 + \sqrt{2 \ln \tfrac{1}{\Theta_{l,2}^{(n)}}} \left( \sum_{u_l \in \mathcal{U}_l } \tau_{\mathcal{P}_{l, u_l,2}^{(n)}}^2 \left( \tfrac{h_{u_l, l, 1}^{(n)}}{h_{l, u_l, 2}^{(n)}} S_{u_l, l}^{(n)} \hat{g}_{l, {u_l^*},  2}^{(n)} \right)^2 \right)^{\tfrac{1}{2}} , ~~ \forall n \hspace{2em} \label{eq:berns-inft-2}
\end{eqnarray}
\end{subequations}

\hrulefill
\vspace*{4pt}
\end{figure*}

\subsection{Sensitivity Analysis}

In the previous section we have seen that the protection functions depend on $\Theta_{l,1}^{(n)}$ and $\Theta_{l,2}^{(n)}$. In the following, we analyze the sensitivity of  $\mathscr{R}_\Delta $ to the values of the trade-off parameters. Using the protections functions  (\ref{eq:berns-inft-1}) and (\ref{eq:berns-inft-2}), $\mathscr{R}_\Delta $ is given by
\begin{equation}
\label{eq:robust_sensitivity}
\mathscr{R}_\Delta \approx \sum_{n=1}^{N} \psi_n^* \tilde{\Delta}_{g_{l, 1}}^{(n)} +  \sum_{n=1}^{N} \varphi_n^* \tilde{\Delta}_{g_{l, 2}}^{(n)} +  \sum_{u_l \in \mathcal{U}_l}\sum_{n=1}^{N} \varrho_{u_l}^{n*} \Delta_{I_{u_l, l}}^{(n)}.
\end{equation}

Differentiating (\ref{eq:robust_sensitivity}) with respect  the to trade-off parameters $\Theta_{l,1}^{(n)}$ and $\Theta_{l,2}^{(n)}$, the sensitivity of $\mathscr{R}_\Delta$, i.e., $\mathcal{S}_{\Theta_{l,i}^{(n)}} \left( \mathscr{R}_\Delta \right) = \frac{\partial \mathscr{R}_\Delta}{\partial \Theta_{l,i}^{(n)}}$ is obtained as follows:
\begin{subequations}
\setlength{\arraycolsep}{0.0em}
\begin{eqnarray}
\hspace{-2em} \mathcal{S}_{\Theta_{l,1}^{(n)}} \left( \mathscr{R}_\Delta \right) & = &  -  \tfrac{ \psi_n^* \left( \displaystyle \sum_{u_l \in \mathcal{U}_l } \tau_{\mathcal{P}_{u_l,l,1}^{(n)}}^2 \left( S_{u_l, l}^{(n)} \hat{g}_{{u_l^*}, l, 1}^{(n)} \right)^2 \right)^{\tfrac{1}{2}}}{\Theta_{l,1}^{(n)} \sqrt{2 \ln \left( \frac{1}{\Theta_{l,1}^{(n)}}\right)}} \\
\hspace{-2em} \mathcal{S}_{\Theta_{l,2}^{(n)}} \left( \mathscr{R}_\Delta \right) & = & - \tfrac{\varphi_n^* \left( \displaystyle \sum_{u_l \in \mathcal{U}_l } \tau_{\mathcal{P}_{l, u_l,2}^{(n)}}^2 \left( \tfrac{h_{u_l, l, 1}^{(n)}}{h_{l, u_l, 2}^{(n)}} S_{u_l, l}^{(n)} \hat{g}_{l, {u_l^*}, 2}^{(n)} \right)^2 \right)^{\tfrac{1}{2}}}{\Theta_{l,2}^{(n)} \sqrt{2 \ln \left( \frac{1}{\Theta_{l,2}^{(n)}}\right)}}. \nonumber \\
\end{eqnarray}
\end{subequations}

\section{Performance Evaluation} \label{sec:performance_eval}

\subsection{Simulation parameters and assumptions}

In order to obtain the performance evaluation results for the proposed resource allocation scheme we use an event-driven simulator in MATLAB. For propagation modeling, we consider distance-dependent path-loss, shadow fading, and multi-path Rayleigh fading. In particular, we consider a realistic 3GPP propagation environment\footnote{Any other propagation model for D2D communication can be used for the proposed resource allocation method.} presented in \cite{relay-book-2}. For example, propagation in UE-to-relay and relay-to-D2D UE links follows the following path-loss equation: $PL_{u_l,l}(\ell)_{[dB]} = 103.8 + 20.9 \log(\ell) + L_{su} + 10 \log(\phi)$, where $\ell$ is the link distance in kilometer; $L_{su}$ accounts for shadow fading and is modelled as a log-normal random variable, and $\phi$ is an exponentially distributed random variable which represents the Rayleigh fading channel power gain. 
For a same link distance, the gains due to shadow fading and Rayleigh fading for different resource blocks could be different.
Similarly, the path-loss equation for relay-eNB link is expressed as $PL_{l,eNB}(\ell)_{[dB]} = 100.7 + 23.5 \log(\ell) + L_{sr} + 10 \log(\phi)$, 
where $L_{sr}$ 
is a log-normal random variable accounting for shadow fading. 
The simulation parameters and assumptions used for obtaining the numerical results are listed in Table \ref{tab:sim_param}. 

\begin{table}[!t]
\renewcommand{\arraystretch}{1.3}
\caption{Simulation Parameters}
\label{tab:sim_param}
\centering
\begin{tabular}{l|l}
\hline
\bfseries Parameter & \bfseries Values\\
\hline\hline
Carrier frequency & $2.35$ GHz \\
System bandwidth & $2.5$ MHz \\
Cell layout & Hexagonal grid, $3$-sector sites \\
Total number of available RBs & $13$ \\
Relay cell radius & $200$ meter\\
Distance between eNB and relays & $125$ meter\\
Minimum distance between UE and relay & $10$ meter\\
Rate requirement for cellular UEs & $128$ Kbps \\
Rate requirement for D2D UEs & $256$ Kbps \\
Total power available at each relay & $30$ dBm \\
Total power available at UE & $23$ dBm \\
Shadow fading standard deviation: \\
 \hspace{5em} for relay-eNB links & $6$ dB \\
 \hspace{5em} for UE-relay links & $10$ dB \\
Noise power spectral density & $-174$ dBm/Hz \\
\hline
\end{tabular}
\end{table}

\begin{figure}[!h t b]
\centering
\includegraphics[width=2.0in]{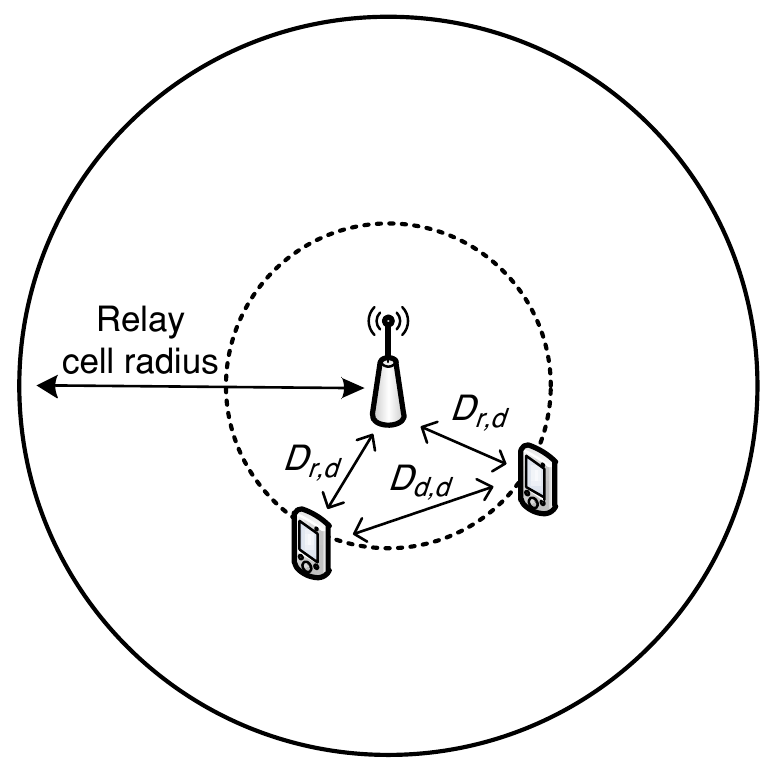}
\caption{Distribution of any D2D-pairs: D2D UEs are uniformly distributed upon the perimeter of circle with radius $D_{r,d}$ and keeping the distance $D_{d,d}$ between peers.} 
\label{fig:d2d_position}
\end{figure}

We simulate a single three-sectored cell in a rectangular
area of $700 ~\text{m} \times 700~ \text{m}$, where the eNB is located in the centre of the cell and three relays are deployed in the network, i.e., one relay in each sector. The CUEs are uniformly distributed within the radius of the relay cell. The D2D transmitters and receivers are uniformly distributed in the perimeter of a circle with radius $D_{r,d}$ as shown in Fig. \ref{fig:d2d_position}. The distance between two D2D UEs is denoted by $D_{d,d}$. Both $D_{r,d}$ and $D_{d,d}$ are varied as  simulation parameters.

In our simulations, we express the uncertainty bounds $\Psi_{l,1}^{(n)}, \Psi_{l,2}^{(n)}$, and $\Upsilon_{u_l}^{(n)}$ in percentage as $\Psi_{l,1}^{(n)} = \frac{{\parallel \mathbf{g}_{l,1}^{(n)} -  \bar{\mathbf{g}}_{l,1}^{(n)} \parallel}_2}{{\parallel \bar{\mathbf{g}}_{l,1}^{(n)} \parallel}_2} $,   $ \Psi_{l,2}^{(n)} = \frac{{\parallel \mathbf{g}_{l,2}^{(n)} -  \bar{\mathbf{g}}_{l,2}^{(n)} \parallel}_2}{{\parallel \bar{\mathbf{g}}_{l,2}^{(n)} \parallel}_2}$, and $\Upsilon_{u_l}^{(n)} = \frac{{\parallel I_{u_l,l}^{(n)} -  \bar{I}_{u_l,l}^{(n)} \parallel}_2}{{\parallel \bar{I}_{u_l,l}^{(n)} \parallel}_2}$. As an example, for any relay node $l$, if $\Psi_{l,1}^{(n)} = 0.5$, the error in the channel gain over RB $n$ for the first hop is not more than $50\%$ of its nominal value. We assume that the estimated interference experienced at relay node and receiving D2D UEs is $\bar{I}_{u_l,l}^{(n)} = 2 \sigma^2$ for all the RBs. The matrices $\mathbf{M}_{g_{l,1}}^{(n)}$ and $\mathbf{M}_{g_{l,2}}^{(n)}$ are considered to be identity matrices and $M_{I_{u_l,l}}^{(n)}$ is set to 1 for all the RBs. The results are obtained by averaging over $250$ realizations of the simulation scenarios (i.e., UE locations and link gains).

\subsection{Results}

\subsubsection{Convergence of the proposed algorithm}

\begin{figure}[!t]
\centering
\includegraphics[width=3.5in]{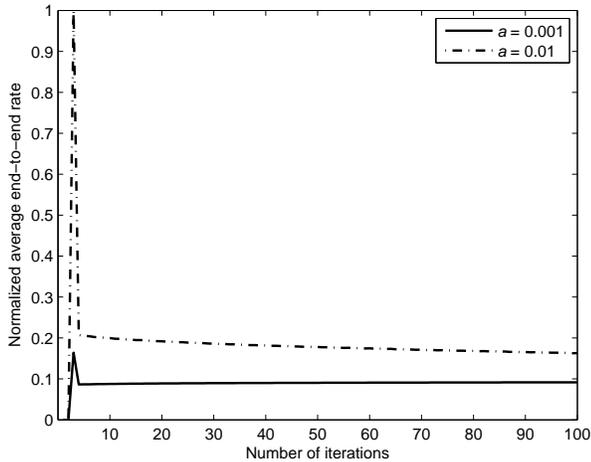}
\caption{Convergence behaviour of the proposed algorithm: number of CUE, $|\mathcal{C}| = 15$ (i.e., $5$ CUEs assisted by each relay), number of D2D pairs,  $|\mathcal{D}| = 9$ (i.e., $3$ D2D pairs are assisted by each relay), and hence $|\mathcal{U}_l| = 8$ for each relay. The average end-to-end-rate is calculated by $\frac{R_l}{|\mathcal{U}_l|}$, the maximum distance between relay-D2D UE, $D_{r,d} = 60~ \text{meter} $, and the interference threshold for both hops is $-70~ \text{dBm}$. The errors (in link gain and experienced interference) are considered to be not more than $50\%$ in each RB.} 
\label{fig:convergence}
\end{figure}

We consider the same step size for all the Lagrange multipliers, i.e., for any Lagrange multiplier $\kappa$, step size at iteration $t$ is calculated as $\Lambda_{\kappa}^{(t)} = \frac{a}{\sqrt{t}}$, where $a$ is a small constant. Fig. \ref{fig:convergence} shows the convergence behavior of the proposed algorithm when $a = 0.001$ and $a = 0.01$. For convergence,  the step size should be selected carefully. It is clear from this figure that when $a$ is sufficiently small, the algorithm converges very quickly (i.e., in less than $20$ iterations) to the optimal solution.

\subsubsection{Sensitivity of $\mathscr{R}_\Delta$ to the trade-off parameter}

\begin{figure}[!t]
\centering
\includegraphics[width=3.5in]{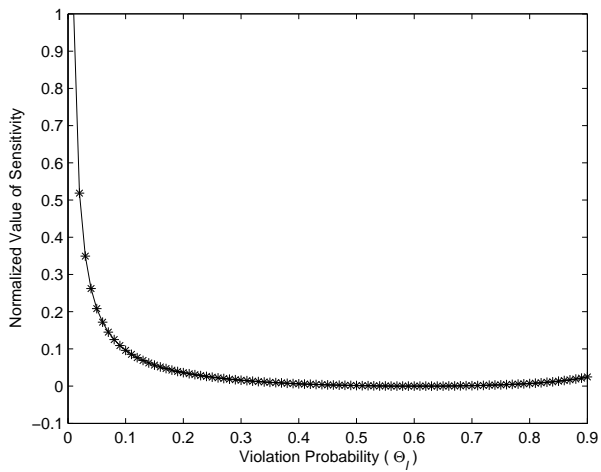}
\caption{Sensitivity of $\mathscr{R}_\Delta$ vs. trade-off parameter using a setup similar to that of Fig. \ref{fig:convergence}. We consider $\hat{g}_{l, 1}^{(n)} = 0.5 \times \bar{g}_{l, 1}^{(n)}$, $\hat{g}_{l, 2}^{(n)} = 0.5  \times \bar{g}_{l, 2}^{(n)}$ and $\Theta_{l,1}^{(n)} = \Theta_{l,2}^{(n)} = \Theta_{l}$ for all the RBs.} 
\label{fig:sensitivity}
\end{figure}

The absolute sensitivity of $\mathscr{R}_\Delta $ considering $\Theta_{l} = \Theta_{l,1}^{(n)} = \Theta_{l,2}^{(n)}$ for $ \forall n$ is shown in Fig.  \ref{fig:sensitivity}. 
For all the RBs, we assume that the probability density function of $\hat{g}_{l, 1}^{(n)}$ and $\hat{g}_{l, 2}^{(n)}$ is Gaussian; hence, $\mathcal{P}_{u_l,l,1}^{(n)}$ and $\mathcal{P}_{l, u_l,2}^{(n)}$ correspond to the last row of Table \ref{tab:berns_val}. For a given uncertainty set and interference threshold, when $\Theta_{l} < 0.2$, the value of $\mathcal{S}_{\Theta_{l}} \left( \mathscr{R}_\Delta \right)$ is very sensitive to $\Theta_{l}$. However, for higher values of $\Theta_{l}$, the sensitivity of $\mathscr{R}_\Delta$ is relatively independent of $\Theta_{l}$. From (\ref{eq:robust_sensitivity}), increasing $\Theta_{l}$ proportionally decreases $\mathscr{R}_\Delta $ which increases network sum-rate. Small values of $\Theta_{l}$ make the system more robust against uncertainty, while higher values of $\Theta_{l}$ increase the network sum-rate. Therefore, by adjusting $\Theta_{l}$ within the range of $0.2$ a trade-off between optimality and robustness can be attained.

\subsubsection{Effect of relaying}

In order to study network performance in presence of the L3 relay, we compare the performance of the proposed scheme with a \textit{reference scheme} \cite{zul-d2d} in which an RB allocated to a CUE can be shared with at most one D2D link. The D2D link shares the same RB(s) (allocated to CUEs using \textbf{Algorithm \ref{alg:rec_alloc}}) and the D2D UEs communicate directly without using the relay only if the QoS requirements for both the CUE and D2D links are satisfied.  

\begin{figure}[!t]
\centering
 \includegraphics[width=3.5in]{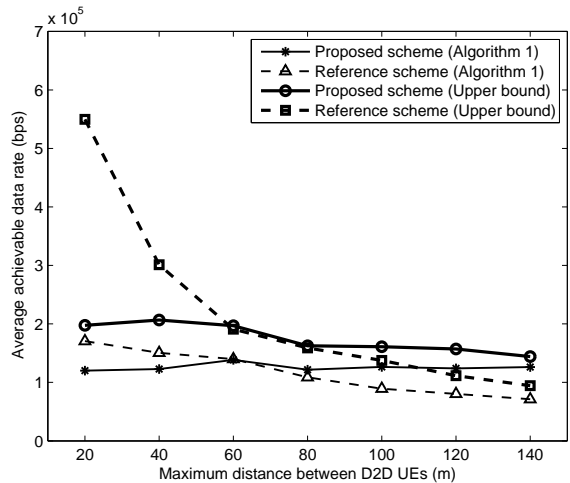}
\caption{Average achievable data rates for D2D UEs in both the proposed and reference schemes compared to the asymptotic upper bound (for $|\mathcal{C}| = 15$, $|\mathcal{D}| = 9$, $D_{r,d}$ = 80 meter and interference threshold = $-70~ \text{dBm}$).} 
\label{fig:gain_5_3_80_ub}
\end{figure}

The average achievable data rate $R_{avg}$ for D2D links is calculated as $R_{avg} =  \frac{ \displaystyle \sum_{u \in \mathcal{D}} R_{u}^{ach}}{|\mathcal{D}|}$, where $R_{u}^{ach}$ is the achievable data rate for link $u$ and $|\cdot|$ denotes the set cardinality. In Fig. \ref{fig:gain_5_3_80_ub}, we compare the performance of \textbf{Algorithm \ref{alg:rec_alloc}} with asymptotic upper bound. Since $\mathbf{P2}$ is a relaxed version of $\mathbf{P1}$, for a sufficiently large number of RBs, the solution obtained by $\mathbf{P2}$ is asymptotically optimal and can be considered as an upper bound \cite{large-rb-dual}. In order to obtain the upper bound, we solve $\mathbf{P2}$ using interior point method (Chapter 11 in \cite{book-boyd}). Note that solving $\mathbf{P2}$ by using the interior point method incurs a complexity of $\mathcal{O}\left( \left( |\mathbf{x}_{\boldsymbol l}| + |\mathbf{S}_{\boldsymbol l}| + |\mathbf{\omega}_{\boldsymbol l}| \right)^3 \right)$ (Chapter 11 in \cite{book-boyd} and \cite{im-complexity}) where $\mathbf{x}_{\boldsymbol l} = \left[ x_{1}^{(1)}, \cdots,  x_{1}^{(N)}, \cdots, x_{|\mathcal{U}_l|}^{(1)}, \cdots, x_{|\mathcal{U}_l|}^{(N)} \right]^\mathsf{T}$, $\mathbf{S}_{\boldsymbol l} = \left[ S_{1,l}^{(1)}, \cdots, S_{1,l}^{(N)}, \cdots,  S_{|\mathcal{U}_l|,l}^{(1)}, \cdots, S_{|\mathcal{U}_l|,l}^{(N)} \right]^\mathsf{T}$ and $\boldsymbol{\omega}_{\boldsymbol l} = \left[ \omega_{1}^{(1)}, \cdots,  \omega_{1}^{(N)}, \cdots, \omega_{|\mathcal{U}_l|}^{(1)}, \cdots, \omega_{|\mathcal{U}_l|}^{(N)} \right]^\mathsf{T}$. From  Fig. \ref{fig:gain_5_3_80_ub} it can be observed that our proposed approach, which uses relays for D2D traffic, can greatly improve the data rate in particular when the distance increases. In addition, proposed algorithm performs close to upper bound with significantly less complexity.  
\begin{figure}[!t]
\centering
 \includegraphics[width=3.5in]{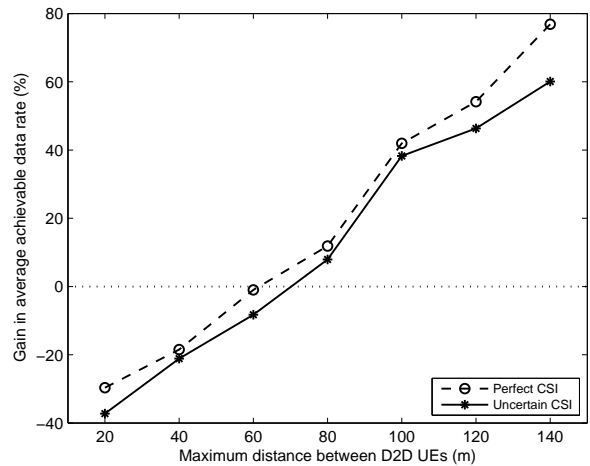}
\caption{Gain in average achievable data rate for D2D UEs (for  $|\mathcal{C}| = 15$, $|\mathcal{D}| = 9$, $D_{r,d}$ = 80 meter and interference threshold = $-70~ \text{dBm}$). For uncertain CSI, the bound on the uncertainty set for channel gain and interference (i.e., $\Psi_{l,1}^{(n)}, \Psi_{l,2}^{(n)}$, and $\Upsilon_{u_l}^{(n)}$) is considered $20\%$ for all the RBs.} 
\label{fig:gain_5_3_80}
\end{figure}

The rate gains for both perfect CSI and under uncertainties are depicted in Fig. \ref{fig:gain_5_3_80}. We calculate the rate gain as follows: $$R_{gain} = \frac{R_{prop} - R_{ref}}{R_{ref}} \times 100 \% $$ where $R_{prop}$ and  $R_{ref}$ denote the average rate for the D2D links in the proposed scheme and the reference scheme, respectively. As expected, under uncertainties, the gain is reduced  compared to the case when perfect channel information is available. Although the reference scheme outperforms when the distance between D2D-link is closer, our proposed approach of relay-aided D2D communication can greatly increase the data rate especially when the distance increases. When the distance between D2D becomes higher, the performance of direct communication deteriorates. Besides, since the D2D links share resources with only one CUE, the spectrum may not be utilized efficiently and this decreases the achievable rate. 

\begin{figure}[!t]
\centering
 \includegraphics[width=3.5in]{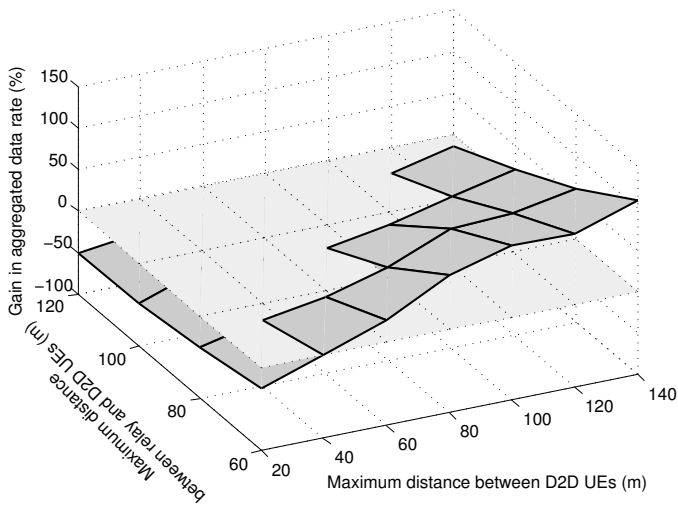}
\caption{Gain in aggregated data rate with different distance between relay and D2D UEs, $D_{r,d}$ where $|\mathcal{C}| = 15$, $|\mathcal{D}| = 9$, interference threshold = $-70~ \text{dBm}$, $\Psi_{l,1}^{(n)}, \Psi_{l,2}^{(n)}$, and $\Upsilon_{u_l}^{(n)}$ are considered $20\%$ for all the RBs. For different values of $D_{r,d}$, there is a distance margin beyond which relaying D2D traffic improves network performance (i.e., the upper portion the of shaded surface where rate gain is positive).} 
\label{fig:gain_5_3_80_rel_dist}
\end{figure}

\begin{figure}[!t]
\centering
\includegraphics[width=3.5in]{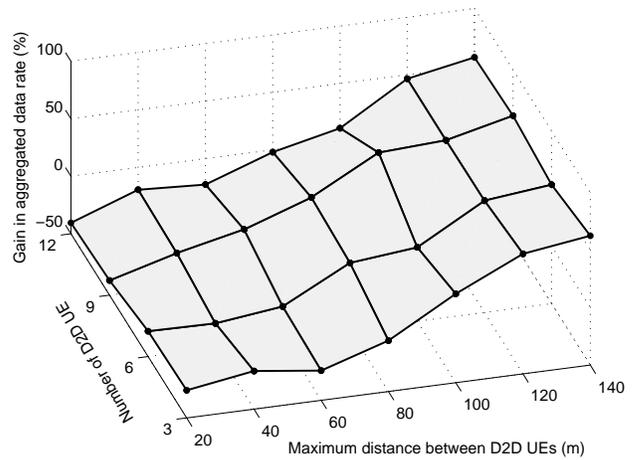}
\caption{Gain in aggregated data rate with varying number of D2D UEs (for $|\mathcal{C}| = 15$, $D_{r,d} = 80~ \text{meter}$, interference threshold = $-70~ \text{dBm}$, $\Psi_{l,1}^{(n)}, \Psi_{l,2}^{(n)}$, and $\Upsilon_{u_l}^{(n)}$ is considered $20\%$ for all the RBs).} 
\label{fig:gain_5_3_80_nd2d}
\end{figure}

The performance gain in terms of the achievable aggregated data rate under different relay-D2D UE distance is shown in Fig. \ref{fig:gain_5_3_80_rel_dist}. It can be observed that, even for relatively large relay-D2D UE distances, e.g., $D_{r,d} \geq 80 ~\text{m}$, relaying D2D traffic provides considerable rate gain for distant D2D UEs. To observe the performance of our proposed scheme in a dense network, we vary the number of D2D UEs and plot the rate gain in Fig. \ref{fig:gain_5_3_80_nd2d}. As can be seen from this figure, even in a moderately dense situation (e.g., $|\mathcal{C}| + |\mathcal{D}| = 15+12 = 27$) our proposed method provides a higher rate compared to that for direct communication between distant D2D UEs.

\section{Conclusion} \label{sec:conclusion}

We have provided a comprehensive resource allocation framework under channel gain uncertainty for relay-assisted D2D communication. Considering two major sources of uncertainty, namely, the link gain between neighbouring relay nodes in both hops and the experienced interference at each receiving network node, the uncertainty has been modeled as a bounded difference between actual and nominal values. By modifying the protection functions in the robust problem, we have shown that the convexity of the problem is maintained. In order to allocate radio resources efficiently, we have proposed a polynomial time distributed algorithm and to balance the cost of robustness defined as the reduction of achievable network sum-rate, we have provided a trade-off mechanism. Through extensive simulations we have observed that, in comparison with a direct D2D communication scheme,  beyond a distance threshold, relaying of D2D traffic for distant D2D UEs significantly improves the network performance. As a future work, this approach can be extended by considering delay as a QoS parameter. Besides, most of the resource allocation problems are formulated under the assumption that the potential D2D UEs have already been discovered. However, to develop a complete D2D communication framework, it is necessary to consider D2D discovery along with resource allocation. 



\appendices
\numberwithin{equation}{section} 

\section{Power and RB Allocation for Nominal Problem} 
\label{app:power-rb-alloc-nominal}

\begin{figure*}[!t]
\normalsize

\begin{flalign}
\boldsymbol{\mathbb{L}}_l (\mathbf{x}, \mathbf{S}, \boldsymbol{\omega}, \boldsymbol{\mu},\boldsymbol{\rho}, \nu_l, \boldsymbol{\psi}, \boldsymbol{\varphi}, \boldsymbol{\lambda}, \boldsymbol{\varrho}) =  \nonumber  
&- \sum_{u_l \in \mathcal{U}_l} \sum_{n = 1}^{N} \tfrac{1}{2} x_{u_l}^{(n)} B_{RB} \log_2 \left(1+ \tfrac{S_{u_l, l}^{(n)} h_{u_l, l, 1}^{(n)}}{x_{u_l}^{(n)} \omega_{u_l}^{(n)}}\right) + \sum_{n = 1}^{N}  \mu_n  \left(\sum_{u_l \in \mathcal{U}_l} x_{u_l}^{(n)} -1 \right) \nonumber \\
&+ \sum_{u_l \in \mathcal{U}_l}  \rho_{u_l} \left( \sum_{n = 1}^{N} S_{u_l, l}^{(n)} - P_{u_l}^{max} \right) + \nu_l \left( \sum_{u_l \in \mathcal{U}_l } \sum_{n =1}^N \tfrac{h_{u_l, l, 1}^{(n)}}{h_{l, u_l, 2}^{(n)}} S_{u_l, l}^{(n)} - P_l^{max} \right) \nonumber \\
&+ \sum_{n = 1}^{N}  \psi_n  \left( \sum_{u_l \in \mathcal{U}_l } S_{u_l, l}^{(n)} g_{{u_l^*}, l, 1}^{(n)} ~ - I_{th, 1}^{(n)} \right) 
+ \sum_{n = 1}^{N}  \varphi_n  \left(\sum_{u_l \in \mathcal{U}_l } \tfrac{h_{u_l, l, 1}^{(n)}}{h_{l, u_l, 2}^{(n)}} S_{u_l, l}^{(n)} g_{l, {u_l^*}, 2}^{(n)} - I_{th, 2}^{(n)} \right) \nonumber \\
&+ \sum_{u_l \in \mathcal{U}_l }  \lambda_{u_l} \left(Q_{u_l} - \sum_{n=1}^N \tfrac{1}{2}  x_{u_l}^{(n)} B_{RB} \log_2 \left(  1 + \tfrac{S_{u_l, l}^{(n)} h_{u_l, l, 1}^{(n)}}{x_{u_l}^{(n)} \omega_{u_l}^{(n)}} \right) \right)  \nonumber \\
&+ \sum_{u_l \in \mathcal{U}_l} \sum_{n = 1}^{N} \varrho_{u_l}^n \left( I_{u_l,l}^{(n)} +  \sigma^2  -  \omega_{u_l}^{(n)} \right).\label{eq:lagrange-1} 
\end{flalign}

\hrulefill
\vspace*{4pt}
\end{figure*}

To observe the nature of power allocation for a UE, we use Karush-Kuhn-Tucker (KKT) optimality conditions and define the Lagrangian function as given in (\ref{eq:lagrange-1}), where $\boldsymbol{\lambda}$ is the vector of Lagrange multipliers associated with individual QoS requirements for cellular and D2D UEs. Similarly, $\boldsymbol{\mu},\boldsymbol{\rho}, \nu_l, \boldsymbol{\psi}, \boldsymbol{\varphi}$ are the Lagrange multipliers for the constraints in (\ref{eq:con-bin-relx})--(\ref{eq:con-intf-2-relx}). Differentiating (\ref{eq:lagrange-1}) with respect to $S_{u_l, l}^{(n)}$, we obtain (\ref{eq:power-alloc}) for power allocation for the link $u_l$ over RB $n$. Similarly, differentiating (\ref{eq:lagrange-1}) with respect to $x_{u_l}^{(n)}$ gives the condition for RB allocation.

\section{Power and RB Allocation for Robust Problem} 
\label{app:power-rb-alloc-robust}

To obtain a more tractable formula, for any vector $\mathbf{y}$ we use the inequality ${\parallel \mathbf{y} \parallel}_2 \leq {\parallel \mathbf{y} \parallel}_1$ and rewrite the constraints (\ref{eq:con-intf-1-robst3}) and (\ref{eq:con-intf-2-robst3}) as (\ref{eq:con-intf-1-robst-mod}) and (\ref{eq:con-intf-2-robst-mod}), respectively, where for any diagonal matrix $\mathbf{A}$, $m_{jj}$ represents the $j$-th element of ${\mathbf{A}}^{-1}(j,:)$. Considering the convexity of $\mathbf{P5}$, the Lagrange dual function can be obtained by (\ref{eq:lagrange-robust}) in which $\boldsymbol{\mu},\boldsymbol{\rho}, \nu_l, \boldsymbol{\psi}, \boldsymbol{\varphi}, \boldsymbol{\lambda}, \boldsymbol{\varrho}$ are the corresponding Lagrange multipliers. Differentiating (\ref{eq:lagrange-robust}) with respect to $S_{u_l, l}^{(n)}$ and $x_{u_l}^{(n)}$ gives (\ref{eq:power-alloc-robust}) and (\ref{eq:channel_alloc1}) for power and RB allocation, respectively.

\begin{figure*}[!htb]
\normalsize

\begin{subequations}
\setlength{\arraycolsep}{0.0em}
\begin{eqnarray}
\sum_{u_l \in \mathcal{U}_l } S_{u_l, l}^{(n)} \bar{g}_{{u_l^*}, l, 1}^{(n)} + \Psi_{l,1}^{(n)}  \sum_{u_l \in \mathcal{U}_l }  m_{{u_l}{u_l}_{g_{l,1}}}^{(n)}  S_{u_l, l}^{(n)}    ~ &&{\leq} ~ I_{th, 1}^{(n)}, ~~ \forall n \label{eq:con-intf-1-robst-mod}\\
\sum_{u_l \in \mathcal{U}_l } \tfrac{h_{u_l, l, 1}^{(n)}}  {h_{l, u_l, 2}^{(n)}} S_{u_l, l}^{(n)} \bar{g}_{l, {u_l^*}, 2}^{(n)} + \Psi_{l,2}^{(n)} \sum_{u_l \in \mathcal{U}_l }  m_{{u_l}{u_l}_{g_{l,2}}}^{(n)}  \tfrac{h_{u_l, l, 1}^{(n)}}{h_{l, u_l, 2}^{(n)}} S_{u_l, l}^{(n)}  ~ &&{\leq} ~ I_{th, 2}^{(n)}, ~~\forall n \label{eq:con-intf-2-robst-mod}
\end{eqnarray}
\setlength{\arraycolsep}{5pt}
\end{subequations}

\hrulefill
\vspace*{4pt}
\end{figure*}

\begin{figure*}[!t]
\normalsize

\begin{flalign}
    {\boldsymbol{\mathbb{L}}_\Delta}_l (\mathbf{x}, \mathbf{S}, \boldsymbol{\omega} \boldsymbol{\mu},\boldsymbol{\rho}, \nu_l, \boldsymbol{\psi}, \boldsymbol{\varphi}, \boldsymbol{\lambda}, \boldsymbol{\varrho}) = &- \sum_{u_l \in \mathcal{U}_l} \sum_{n = 1}^{N} \tfrac{1}{2} x_{u_l}^{(n)} B_{RB} \log_2 \left(1+ \tfrac{S_{u_l, l}^{(n)} h_{u_l, l, 1}^{(n)}}{x_{u_l}^{(n)} \omega_{u_l}^{(n)}}\right) + \sum_{n = 1}^{N}  \mu_n  \left(\sum_{u_l \in \mathcal{U}_l} x_{u_l}^{(n)} -1 \right) \nonumber\\  
&+ \sum_{u_l \in \mathcal{U}_l}  \rho_{u_l} \left( \sum_{n = 1}^{N} S_{u_l, l}^{(n)} - P_{u_l}^{max} \right) + \nu_l \left( \sum_{u_l \in \mathcal{U}_l } \sum_{n =1}^N \tfrac{h_{u_l, l, 1}^{(n)}}{h_{l, u_l, 2}^{(n)}} S_{u_l, l}^{(n)} - P_l^{max} \right) \nonumber \\
&+ \sum_{n = 1}^{N}  \psi_n  \left(   \sum_{u_l \in \mathcal{U}_l } S_{u_l, l}^{(n)} \bar{g}_{{u_l^*}, l, 1}^{(n)}  +  \Psi_{l,1}^{(n)}  \sum_{u_l=1}^{|\mathcal{U}_l|} \left( m_{{u_l}{u_l}_{g_{l,1}}}^{(n)}  S_{u_l, l}^{(n)}  \right) - I_{th, 1}^{(n)} \right) \nonumber \\
&+ \sum_{n = 1}^{N}  \varphi_n  \left(\sum_{u_l \in \mathcal{U}_l } \tfrac{h_{u_l, l, 1}^{(n)}}{h_{l, u_l, 2}^{(n)}} S_{u_l, l}^{(n)} \bar{g}_{l, {u_l^*}, 2}^{(n)} + \Psi_{l,2}^{(n)} \sum_{u_l=1}^{|\mathcal{U}_l|} \left( m_{{u_l}{u_l}_{g_{l,2}}}^{(n)}  \tfrac{h_{u_l, l, 1}^{(n)}}{h_{l, u_l, 2}^{(n)}} S_{u_l, l}^{(n)}  \right) - I_{th, 2}^{(n)} \right) \nonumber \\
&+ \sum_{u_l \in \mathcal{U}_l }  \lambda_{u_l} \left(Q_{u_l} - \sum_{n=1}^N \tfrac{1}{2}  x_{u_l}^{(n)} B_{RB} \log_2 \left(  1 + \tfrac{S_{u_l, l}^{(n)} \gamma_{u_l, l, 1}^{(n)}}{x_{u_l}^{(n)} \omega_{u_l}^{(n)}} \right) \right)  \nonumber \\
&+ \sum_{u_l \in \mathcal{U}_l} \sum_{n = 1}^{N} \varrho_{u_l}^n \left( \bar{I}_{u_l,l}^{(n)} + \Delta_{I_{u_l,l}}^{(n)} + \sigma^2  -  \omega_{u_l}^{(n)} \right).
\label{eq:lagrange-robust} 
\end{flalign}

\hrulefill
\vspace*{4pt}
\end{figure*}

\section{Update of variables and Lagrange multipliers} \label{app:lagrange_update}

After finding the optimal solution, i.e., ${P_{u_l,l}^{(n)}}^*$ and ${x_{u_l}^{(n)}}^*$, the primal and dual variables at the $(t+1)$-th iteration are updated using (\ref{eq:lagrange_update1})--(\ref{eq:lagrange_update8}), where $\Lambda_{\kappa}^{(t)}$ is the small step size for variable $\kappa$ at iteration $t$ and the partial derivative of the Lagrange dual function with respect to $\omega_{u_l}^{(n)}$ is
\begin{equation}
\label{eq:partial_omega}
\frac{\partial {\boldsymbol{\mathbb{L}}_\Delta}_l}{\partial \omega_{u_l}^{(n)}} = 
 \tfrac{1}{2} B_{RB} \frac{\left(\lambda_{u_l} + 1 \right) x_{u_l}^{(n)} S_{u_l,l}^{(n)} h_{u_l,l,1}^{(n)}}{ \omega_{u_l}^{(n)} \left(x_{u_l}^{(n)}\omega_{u_l}^{(n)} + S_{u_l,l}^{(n)} h_{u_l,l,1}^{(n)} \right) \ln 2} - \varrho_{u_l}^n.
\end{equation}

\begin{figure*}[!t]
\normalsize

\begin{subequations}
\setlength{\arraycolsep}{0.0em}

\begin{align}
\omega_{u_l}^{(n)}(t+1) &= \left[ \omega_{u_l}^{(n)}(t) - \Lambda_{\omega_{u_l}^{(n)}}^{(t)} \frac{\partial \boldsymbol{\mathbb{L}_l}}{\partial \omega_{u_l}^{(n)}}\bigg|_{t} \right]^+ \label{eq:lagrange_update1}\\
 \mu_n(t+1) &=  \left[ \mu_n(t) + \Lambda_{\mu_n}^{(t)} \left(\sum_{u_l \in \mathcal{U}_l} x_{u_l}^{(n)} -1 \right)  \right]^+ \label{eq:lagrange_update2}\\
 \rho_{u_l}(t+1) &= \left[ \rho_{u_l}(t) + \Lambda_{\rho_{u_l}}^{(t)} \left( \sum_{n = 1}^{N} S_{u_l, l}^{(n)} - P_{u_l}^{max} \right) \right]^+ \label{eq:lagrange_update3}\\
 \nu_l(t+1) &= \left[ \nu_l(t) + \Lambda_{\nu_l}^{(t)} \left( \sum_{u_l \in \mathcal{U}_l } \sum_{n =1}^N \tfrac{h_{u_l, l, 1}^{(n)}}{h_{l, u_l, 2}^{(n)}} S_{u_l, l}^{(n)} - P_l^{max} \right) \right]^+ \label{eq:lagrange_update4}\\
 \psi_n(t+1) &= \left[ \psi_n(t) + \Lambda_{\psi_n}^{(t)} \left(   \sum_{u_l \in \mathcal{U}_l } S_{u_l, l}^{(n)} \bar{g}_{{u_l^*}, l, 1}^{(n)}  +  \Psi_{l,1}^{(n)}  \sum_{u_l=1}^{|\mathcal{U}_l|} \left( m_{{u_l}{u_l}_{g_{l,1}}}^{(n)}  S_{u_l, l}^{(n)}  \right) - I_{th, 1}^{(n)} \right) \right]^+ \label{eq:lagrange_update5}\\
 \varphi_n(t+1) &= \left[ \varphi_n(t) + \Lambda_{\varphi_n}^{(t)} \left(\sum_{u_l \in \mathcal{U}_l } \tfrac{h_{u_l, l, 1}^{(n)}}{h_{l, u_l, 2}^{(n)}} S_{u_l, l}^{(n)} \bar{g}_{l, {u_l^*}, 2}^{(n)} + \Psi_{l,2}^{(n)} \sum_{u_l=1}^{|\mathcal{U}_l|} \left( m_{{u_l}{u_l}_{g_{l,2}}}^{(n)}  \tfrac{h_{u_l, l, 1}^{(n)}}{h_{l, u_l, 2}^{(n)}} S_{u_l, l}^{(n)}  \right) - I_{th, 2}^{(n)} \right)  
  \right]^+ \label{eq:lagrange_update6}\\
  \lambda_{u_l}(t+1) &= \left[ \lambda_{u_l}(t) + \Lambda_{\lambda_{u_l}}^{(t)} \left(Q_{u_l} - \sum_{n=1}^N \tfrac{1}{2}  x_{u_l}^{(n)} B_{RB} \log_2 \left(  1 + \tfrac{S_{u_l, l}^{(n)} \gamma_{u_l, l, 1}^{(n)}}{x_{u_l}^{(n)} \omega_{u_l}^{(n)}} \right) \right)
    \right]^+ \label{eq:lagrange_update7} \\
\varrho_{u_l}^n(t+1) &= \left[ \varrho_{u_l}^n(t) + \Lambda_{\varrho_{u_l}^n}^{(t)}  \left( \bar{I}_{u_l,l}^{(n)} + \Delta_{I_{u_l,l}}^{(n)} + \sigma^2  -  \omega_{u_l}^{(n)} \right)
    \right]^+.  \label{eq:lagrange_update8}      
\end{align}

\setlength{\arraycolsep}{5pt}
\end{subequations}

\hrulefill
\vspace*{4pt}
\end{figure*}


\section{Proof of Proposition \ref{theorem:robust-tradeoff}}
\label{app:sensitivity}

\begin{figure*}[!t]
\normalsize

\begin{eqnarray}
\label{eq:inf_function}
&\mathscr{R}^*(\mathbf{a}, \mathbf{b}, \mathbf{c}) = \inf \Bigg\{ \left.  \underset{x_{u_l}^{(n)}, S_{u_l, l}^{(n)}, \omega_{u_l}^{(n)}}{\operatorname{max}} ~ \displaystyle \sum_{u_l \in \mathcal{U}_l }  \sum_{n =1}^N   \frac{1}{2}  x_{u_l}^{(n)}   B_{RB}  \log_2 \left(  1 +   \frac{S_{u_l, l}^{(n)} h_{u_l, l, 1}^{(n)}}{x_{u_l}^{(n)} \omega_{u_l}^{(n)}} \right)   \right\vert, \nonumber \\ 
& \displaystyle \sum_{u_l \in \mathcal{U}_l} x_{u_l}^{(n)}  \leq  1, \quad \sum_{n =1}^N S_{u_l, l}^{(n)}  \leq  P_{u_l}^{max}, \quad \sum_{u_l \in \mathcal{U}_l } \sum_{n =1}^N \frac{h_{u_l, l, 1}^{(n)}}{h_{l, u_l, 2}^{(n)}} S_{u_l, l}^{(n)} \leq P_l^{max}, \nonumber\\
& \displaystyle \sum_{u_l \in \mathcal{U}_l } S_{u_l, l}^{(n)} \bar{g}_{{u_l^*}, l, 1}^{(n)} + \Delta_{g_{l, 1}}^{(n)} \leq I_{th, 1}^{(n)}, \quad \sum_{u_l \in \mathcal{U}_l } \frac{h_{u_l, l, 1}^{(n)}}  {h_{l, u_l, 2}^{(n)}} S_{u_l, l}^{(n)} \bar{g}_{l, {u_l^*},  2}^{(n)} + \Delta_{g_{l, 2}}^{(n)}  \leq I_{th, 2}^{(n)},  \nonumber\\
& \displaystyle \sum_{n=1}^N \frac{1}{2}  x_{u_l}^{(n)} B_{RB} \log_2 \left(  1 + \frac{S_{u_l, l}^{(n)} h_{u_l, l, 1}^{(n)}}{x_{u_l}^{(n)} \omega_{u_l}^{(n)}} \right)  \geq  Q_{u_l}, \quad
S_{u_l, l}^{(n)}  \geq 0, \quad
\bar{I}_{u_l,l}^{(n)} + \Delta_{I_{u_l,l}}^{(n)} + \sigma^2 \leq \omega_{u_l}^{(n)}  \Bigg\}.
\end{eqnarray}

\hrulefill
\vspace*{4pt}
\end{figure*}

Since $\mathbf{P4}$ is a perturbed version of $\mathbf{P2}$ with protection functions in the constraints (\ref{eq:con-intf-1-relx}), (\ref{eq:con-intf-2-relx}), and (\ref{eq:con-aux-relx}), to obtain (\ref{eq:robust-opt-tradeoff}), we use local sensitivity analysis of $\mathbf{P4}$ by perturbing its constraints (Chapter IV in \cite{sensitivity_book}, Section 5.6 in \cite{book-boyd}). Let the elements of $\mathbf{a}, \mathbf{b}, \mathbf{c}$ contain $\Delta_{g_{l, 1}}^{(n)}, \Delta_{g_{l, 2}}^{(n)}$ $\forall n$, and $\Delta_{I_{u_l, l}}^{(n)}$ $\forall u_l, n$, where $\mathscr{R}^*(\mathbf{a}, \mathbf{b}, \mathbf{c})$ is given by (\ref{eq:inf_function}). When $\Delta_{g_{l, 1}}^{(n)}, \Delta_{g_{l, 2}}^{(n)}$, and $\Delta_{I_{u_l, l}}^{(n)}$ are small, $\mathscr{R}^*(\mathbf{a}, \mathbf{b}, \mathbf{c})$ is differentiable with respect to the perturbation vectors $\mathbf{a}, \mathbf{b}$, and $ \mathbf{c}$ (Chapter IV in \cite{sensitivity_book}). Using Taylor series, (\ref{eq:inf_function}) can be written as
\begin{eqnarray}
\label{eq:taylor}
\mathscr{R}^*(\mathbf{a}, \mathbf{b}, \mathbf{c}) = \mathscr{R}^*(\mathbf{0}, \mathbf{0}, \mathbf{0}) + 
\sum_{n=1}^{N} a_n \frac{\partial \mathscr{R}^*(\mathbf{0}, \mathbf{b}, \mathbf{c})}{\partial a_n} + \nonumber \\
  \sum_{n=1}^{N} b_n \frac{\partial \mathscr{R}^*(\mathbf{a}, \mathbf{0}, \mathbf{c})}{\partial b_n}  +  \sum_{u_l \in \mathcal{U}_l}\sum_{n=1}^{N} c_{u_l}^n \frac{\partial \mathscr{R}^*(\mathbf{a}, \mathbf{b}, \mathbf{c})}{\partial c_{u_l}^n} + o \nonumber \\
\end{eqnarray}
where $\mathscr{R}^*(\mathbf{0}, \mathbf{0}, \mathbf{0}) $ is the optimal value for $\mathbf{P2}$, $\mathbf{0}$ is the zero vector, and $o$ is the truncation error in the Taylor series expansion. Note that 
$\mathscr{R}^*(\mathbf{0}, \mathbf{0}, \mathbf{0})$ and $\mathscr{R}^*(\mathbf{a}, \mathbf{b}, \mathbf{c})$ are equal to $R^*$ and $R_\Delta^*$, respectively. Since $\mathbf{P2}$ is convex, $\mathscr{R}^*(\mathbf{a}, \mathbf{b}, \mathbf{c})$ is obtained from the Lagrange dual function [i.e., (\ref{eq:lagrange-1})] of $\mathbf{P2}$; and using the sensitivity analysis (Chapter IV in \cite{sensitivity_book}), we have $\frac{\partial \mathscr{R}^*(\mathbf{0}, \mathbf{b}, \mathbf{c})}{\partial a_n} \approx -\psi_n^*$, $\frac{\partial \mathscr{R}^*(\mathbf{a}, \mathbf{0}, \mathbf{c})}{\partial b_n} \approx -\varphi_n^*$ and $\frac{\partial \mathscr{R}^*(\mathbf{a}, \mathbf{b}, \mathbf{0})}{\partial  c_{u_l}^n} \approx -\varrho_{u_l}^{n*}$. Rearranging (\ref{eq:taylor}) we obtain
\begin{equation}
\label{eq:sensitivity-proof}
R_\Delta^* - R^* \approx - \sum_{n=1}^{N} \psi_n^* \Delta_{g_{l, 1}}^{(n)} -  \sum_{n=1}^{N} \varphi_n^* \Delta_{g_{l, 2}}^{(n)} -  \sum_{u_l \in \mathcal{U}_l}\sum_{n=1}^{N} \varrho_{u_l}^{n*} \Delta_{I_{u_l, l}}^{(n)}.
\end{equation}

Since $\psi_n^*, \varphi_n^*, \varrho_{u_l}^{n*} $ are non-negative Lagrange multipliers, the achievable sum-rate is reduced compared to the case in which perfect channel information is available.

\section{Parameters used for approximations in the chance constraint approach}
\label{app:berns_val}

In order to balance the robustness and optimality, the parameters used for safe approximations of the chance constraints (obtained from \cite{robust-theroy}) are given in Table \ref{tab:berns_val}.

\begin{table}[!h]
\renewcommand{\arraystretch}{1.3}
\caption{Values of $\eta_{\mathcal{P}_{j}}^{+}$ and $\tau_{\mathcal{P}_{j}}$ for Typical Families of Probability Distribution $\mathcal{P}_{j}$}
\label{tab:berns_val}
\centering
\begin{tabular}{l||c|c}
\hline
 $\mathcal{P}_{j}$ &  $\eta_{\mathcal{P}_{j}}^{+}$ & $\tau_{\mathcal{P}_{j}}$\\
\hline\hline
 $\sup \left\lbrace \mathcal{P}_{j} \right\rbrace \in [-1,+1]$ & $1$ & $0$ \\
 $\sup \left\lbrace \mathcal{P}_{j} \right\rbrace$ is unimodal and $\sup \left\lbrace \mathcal{P}_{j} \right\rbrace \in [-1,+1]$ & $\frac{1}{2}$ & $\frac{1}{\sqrt{12}}$ \\
 $\sup \left\lbrace \mathcal{P}_{j} \right\rbrace$ is unimodal and symmetric & $0$ & $\frac{1}{\sqrt{3}}$ \\
\hline
\end{tabular}
\end{table}

\section*{Acknowledgment}
This work was supported in part by a University of Manitoba Graduate Fellowship, in part by the Natural Sciences and Engineering Research Council of Canada (NSERC) Strategic  Grant (STPGP 430285), and in part by the National Research Foundation of Korea (NRF) grant funded by the Korean government (MSIP) (NRF-2013R1A2A2A01067195).


\bibliographystyle{IEEEtran}

\begin{thebibliography}{10}
\providecommand{\url}[1]{#1}
\csname url@samestyle\endcsname
\providecommand{\newblock}{\relax}
\providecommand{\bibinfo}[2]{#2}
\providecommand{\BIBentrySTDinterwordspacing}{\spaceskip=0pt\relax}
\providecommand{\BIBentryALTinterwordstretchfactor}{4}
\providecommand{\BIBentryALTinterwordspacing}{\spaceskip=\fontdimen2\font plus
\BIBentryALTinterwordstretchfactor\fontdimen3\font minus
  \fontdimen4\font\relax}
\providecommand{\BIBforeignlanguage}[2]{{%
\expandafter\ifx\csname l@#1\endcsname\relax
\typeout{** WARNING: IEEEtran.bst: No hyphenation pattern has been}%
\typeout{** loaded for the language `#1'. Using the pattern for}%
\typeout{** the default language instead.}%
\else
\language=\csname l@#1\endcsname
\fi
#2}}
\providecommand{\BIBdecl}{\relax}
\BIBdecl

\bibitem{d2d_example}
L.~Lei, Z.~Zhong, C.~Lin, and X.~Shen, ``{Operator controlled device-to-device
  communications in LTE-Advanced networks},'' \emph{IEEE Wireless
  Communications}, vol.~19, no.~3, pp. 96--104, 2012.

\bibitem{d2d_example2}
M.~Corson, R.~Laroia, J.~Li, V.~Park, T.~Richardson, and G.~Tsirtsis, ``Toward
  proximity-aware internetworking,'' \emph{IEEE Wireless Communications},
  vol.~17, no.~6, pp. 26--33, 2010.

\bibitem{3gpp:d2d_example}
``{Technical specification group services and system aspects; Feasibility study
  for proximity services (ProSe), release 12},'' 3rd Generation Partnership
  Project, Tech. Rep. 3GPP TR 22.803 V12.2.0, June 2013.

\bibitem{phond-d2d}
P.~Phunchongharn, E.~Hossain, and D.~Kim, ``{Resource allocation for
  device-to-device communications underlaying LTE-advanced networks},''
  \emph{IEEE Wireless Communications}, vol.~20, no.~4, pp. 91--100, 2013.

\bibitem{d2d_swarm}
L.~Su, Y.~Ji, P.~Wang, and F.~Liu, ``{Resource allocation using particle swarm
  optimization for D2D communication underlay of cellular networks},'' in Proc. of
  \emph{IEEE Wireless Communications and Networking Conference (WCNC)}, 2013,
  pp. 129--133.

\bibitem{d2d_m2m_1}
N.~K. Pratas and P.~Popovski, ``{Low-rate machine-type communication via
  wireless device-to-device (D2D) links},'' submitted to \emph{IEEE Journal
  on Selected Areas in Communications}, Special Issue on
  ``Device-to-Device Communications in Cellular Networks'', 2013, arXiv
  preprint, \url{http://arxiv.org/abs/1305.6783}.

\bibitem{m2m_our_paper}
M.~Hasan, E.~Hossain, and D.~Niyato, ``{Random access for machine-to-machine
  communication in LTE-advanced networks: issues and approaches},'' \emph{IEEE
  Communications Magazine}, vol.~51, no.~6, pp. 86--93, 2013.

\bibitem{zul-d2d}
M.~Zulhasnine, C.~Huang, and A.~Srinivasan, ``{Efficient resource allocation
  for device-to-device communication underlaying LTE network},'' in Proc. of \emph{IEEE
  6th International Conference on Wireless and Mobile Computing, Networking and
  Communications (WiMob)}, 2010, pp. 368--375.

\bibitem{d2d_new_paper}
Y.~Pei and Y.-C. Liang, ``Resource allocation for device-to-device
  communications overlaying two-way cellular networks,'' \emph{IEEE
  Transactions on Wireless Communications}, vol.~12, no.~7, pp. 3611--3621,
  2013.

\bibitem{lingyang-icc13}
R.~Zhang, L.~Song, Z.~Han, X.~Cheng, and B.~Jiao, ``Distributed resource
  allocation for device-to-device communications underlaying cellular
  networks,'' in Proc. of \emph{IEEE International Conference on Communications (ICC)},
  2013, pp. 1889--1893.

\bibitem{xen-1}
M.~Alam, J.~W. Mark, and X.~Shen, ``{Relay selection and resource allocation
  for multi-user cooperative LTE-A uplink},'' in Proc. of \emph{IEEE International
  Conference on Communications (ICC)}, 2012, pp. 5092--5096.

\bibitem{d2d_first_relay}
Y.-D. Lin and Y.-C. Hsu, ``Multihop cellular: a new architecture for wireless
  communications,'' in Proc. of \emph{IEEE INFOCOM}, vol.~3, 2000, pp. 1273--1282.

\bibitem{d2d-rel-4}
X.~Ma, R.~Yin, G.~Yu, and Z.~Zhang, ``A distributed relay selection method for
  relay assisted device-to-device communication system,'' in Proc. of \emph{IEEE 23rd
  International Symposium on Personal Indoor and Mobile Radio Communications
  (PIMRC)}, 2012, pp. 1020--1024.

\bibitem{d2d-rel-3}
L.~Wang, T.~Peng, Y.~Yang, and W.~Wang, ``{Interference constrained relay
  selection of D2D communication for relay purpose underlaying cellular
  networks},'' in Proc. of \emph{8th International Conference on Wireless
  Communications, Networking and Mobile Computing (WiCOM)}, 2012, pp. 1--5.

\bibitem{d2d-rel-1}
D.~Lee, S.-I. Kim, J.~Lee, and J.~Heo, ``{Performance of multihop
  decode-and-forward relaying assisted device-to-device communication
  underlaying cellular networks},'' in Proc. of  \emph{International Symposium on
  Information Theory and its Applications (ISITA)}, 2012, pp. 455--459.

\bibitem{d2d_relay_2}
K.~Vanganuru, S.~Ferrante, and G.~Sternberg, ``{System capacity and coverage of
  a cellular network with D2D mobile relays},'' in Proc. of  \emph{IEEE Military
  Communications Conference (MILCOM)}, 2012, pp. 1--6.

\bibitem{book-boyd}
S.~Boyd and L.~Vandenberghe, \emph{{Convex Optimization}}. 2004.

\bibitem{worst-case_robust}
A.~Ben-tal and A.~Nemirovski, ``Robust solutions of uncertain linear
  programs,'' \emph{Operations Research Letters}, vol.~25, pp. 1--13, 1999.

\bibitem{bayesian_robust_1}
X.~Zhang, D.~Palomar, and B.~Ottersten, ``{Statistically robust design of
  linear MIMO transceivers},'' \emph{IEEE Transactions on Signal Processing},
  vol.~56, no.~8, pp. 3678--3689, 2008.

\bibitem{bayesian_robust_2}
S.~Zhou and G.~Giannakis, ``Optimal transmitter eigen-beamforming and
  space-time block coding based on channel mean feedback,'' \emph{IEEE
  Transactions on Signal Processing}, vol.~50, no.~10, pp. 2599--2613, 2002.

\bibitem{uncertainity-early}
M.~M\'edard, ``The effect upon channel capacity in wireless communications of
  perfect and imperfect knowledge of the channel.'' \emph{IEEE Transactions on
  Information Theory}, vol.~46, no.~3, pp. 933--946, 2000.

\bibitem{uncertainity-cr-2}
S.-J. Kim, N.~Soltani, and G.~Giannakis, ``{Resource allocation for OFDMA
  cognitive radios under channel uncertainty},'' \emph{IEEE Transactions on
  Wireless Communications}, vol.~12, no.~7, pp. 3578--3587, 2013.

\bibitem{robust_rel}
S.~Mallick, R.~Devarajan, M.~Rashid, and V.~Bhargava, ``Robust power allocation
  designs for cognitive radio networks with cooperative relays,'' in Proc. of \emph{IEEE
  International Conference on Communications (ICC)}, 2012, pp. 1677--1682.

\bibitem{robust_pow}
S.~Parsaeefard and A.~R. Sharafat, ``Robust distributed power control in
  cognitive radio networks,'' \emph{IEEE Transactions on Mobile Computing},
  vol.~12, no.~4, pp. 609--620, 2013.

\bibitem{robust_intf}
S.~Parsaeefard and A.-R. Sharafat, ``Robust worst-case interference control in
  underlay cognitive radio networks,'' \emph{IEEE Transactions on Vehicular
  Technology}, vol.~61, no.~8, pp. 3731--3745, 2012.

\bibitem{d2d_our_paper}
M.~Hasan and E.~Hossain, ``Resource allocation for network-integrated
  device-to-device communications using smart relays,'' in Proc. of \emph{IEEE Global
  Communications Conference (GLOBCOM), Workshop on ``Device-to-Device (D2D)
  Communication with and without Infrastructure}, 2013.

\bibitem{relay-book-1}
D. I.~Kim, W.~Choi, and B.-H. Kim, ``{Partial information relaying and
  relaying in 3GPP LTE},'' in \emph{{Cooperative Cellular Wireless
  Networks}}.\hskip 1em plus 0.5em minus 0.4em\relax Cambridge University
  Press, 2011.

\bibitem{mutihop-rate}
M.~Sikora, J.~Laneman, M.~Haenggi, D.~Costello, and T.~Fuja, ``Bandwidth- and
  power-efficient routing in linear wireless networks,'' \emph{IEEE
  Transactions on Information Theory}, vol.~52, no.~6, pp. 2624--2633, 2006.

\bibitem{ref_user}
K.~Son, S.~Lee, Y.~Yi, and S.~Chong, ``{REFIM: A practical interference
  management in heterogeneous wireless access networks},'' \emph{IEEE Journal
  on Selected Areas in Communications}, vol.~29, no.~6, pp. 1260--1272, 2011.

\bibitem{relax-con-1}
Z.~Shen, J.~Andrews, and B.~Evans, ``{Adaptive resource allocation in multiuser
  OFDM systems with proportional rate constraints},'' \emph{IEEE Transactions
  on Wireless Communications}, vol.~4, no.~6, pp. 2726--2737, 2005.

\bibitem{time-share-1}
M.~Tao, Y.-C. Liang, and F.~Zhang, ``{Resource allocation for delay
  differentiated traffic in multiuser OFDM systems},'' \emph{IEEE Transactions
  on Wireless Communications}, vol.~7, no.~6, pp. 2190--2201, 2008.

\bibitem{large-rb-dual}
W.~Yu and R.~Lui, ``{Dual methods for nonconvex spectrum optimization of
  multicarrier systems},'' \emph{IEEE Transactions on Communications}, vol.~54,
  no.~7, pp. 1310--1322, 2006.

\bibitem{robust-theroy}
A.~Ben-Tal and A.~Nemirovski, ``\BIBforeignlanguage{English}{{Selected topics
  in robust convex optimization}},''
  \emph{\BIBforeignlanguage{English}{Mathematical Programming}}, vol. 112,
  no.~1, pp. 125--158, 2008.

\bibitem{general_norm}
D.~Bertsimas, D.~Pachamanova, and M.~Sim, ``{Robust linear optimization under
  general norms},'' \emph{Operations Research Letters}, vol.~32, no.~6, pp.
  510--516, Nov. 2004.

\bibitem{book:fading_uncor}
K.~Fazel and S.~Kaiser, \emph{Multi-Carrier and Spread Spectrum Systems}.\hskip
  1em plus 0.5em minus 0.4em\relax New York, NY, USA: John Wiley \& Sons, Inc.,
  2003.

\bibitem{fading_uncor2}
K.~Son, B.~C. Jung, S.~Chong, and D.~K. Sung, ``{Power allocation for
  OFDM-based cognitive radio systems under outage constraints},'' in Proc. of \emph{IEEE
  International Conference on Communications (ICC)}, 2010, pp. 1--5.

\bibitem{ellip-m2}
A.~J.~G. Anandkumar, A.~Anandkumar, S.~Lambotharan, and J.~Chambers, ``Robust
  rate maximization game under bounded channel uncertainty,'' \emph{IEEE
  Transactions on Vehicular Technology}, vol.~60, no.~9, pp. 4471--4486, 2011.

\bibitem{ellip-m2-1}
A.~Pascual-Iserte, D.~Palomar, A.~Perez-Neira, and M.-A. Lagunas, ``{A robust
  maximin approach for MIMO communications with imperfect channel state
  information based on convex optimization},'' \emph{IEEE Transactions on
  Signal Processing}, vol.~54, no.~1, pp. 346--360, 2006.

\bibitem{book:conic}
A.~Ben-Tal and A.~S. Nemirovskiaei, ``{Conic quadratic programming},'' in
  \emph{{Lectures on Modern Convex Optimization: Analysis, Algorithms, and
  Engineering Applications (MPS-SIAM Series on Optimization)}}.\hskip 1em plus
  0.5em minus 0.4em\relax Philadelphia, PA, USA: Society for Industrial and
  Applied Mathematics, 2001.

\bibitem{lte_arch}
{Alcatel-Lucent}, ``{The LTE network architecture},'' December 2009, {White
  Paper}.

\bibitem{notes_subgrad}
S.~Boyd and A.~Mutapcic, ``{Subgradient methods},'' [Online],
  \url{www.stanford.edu/class/ee364b/notes/subgrad_method_notes.pdf}, lecture
  notes for EE364b, Stanford University.

\bibitem{relay-book-2}
Y.~Yuan, ``{LTE-A relay scenarios and evaluation methodology},'' in
  \emph{LTE-Advanced Relay Technology and Standardization}.\hskip 1em plus
  0.5em minus 0.4em\relax Springer, 2013, pp. 9--38.

\bibitem{im-complexity}
D.~Bharadia, G.~Bansal, P.~Kaligineedi, and V.~K.~Bhargava, ``{Relay and power
  allocation schemes for OFDM-based cognitive radio systems},'' \emph{IEEE
  Transactions on Wireless Communications}, vol.~10, no.~9, pp. 2812--2817,
  2011.

\bibitem{sensitivity_book}
D.~G. Cacuci, \emph{{Sensitivity and Uncertainty Analysis. Vol. 1:
  Theory}}.\hskip 1em plus 0.5em minus 0.4em\relax Boca Raton, FL: Chapman and
  Hall/CRC, 2003.

\end{thebibliography}


\begin{IEEEbiography} [{\includegraphics[width=1in,height=1.25in,clip,keepaspectratio]{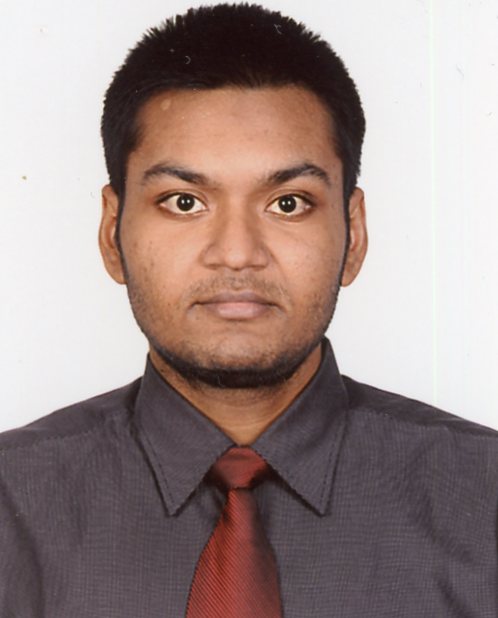}}]
{Monowar Hasan} (S'13)  received his B.Sc. degree in Computer Science and Engineering from Bangladesh University of Engineering and Technology (BUET), Dhaka, in 2012. He is currently working toward his M.Sc. degree in the Department of Electrical and Computer Engineering at the University of Manitoba, Winnipeg, Canada. He has been awarded the University of Manitoba Graduate Fellowship. His current research interests include internet of things, nano-communication networks and resource allocation in mobile cloud computing. He has served as a reviewer for several major IEEE magazines and journals.

\end{IEEEbiography}

\begin{IEEEbiography} [{\includegraphics[width=1in,height=1.25in,clip,keepaspectratio]{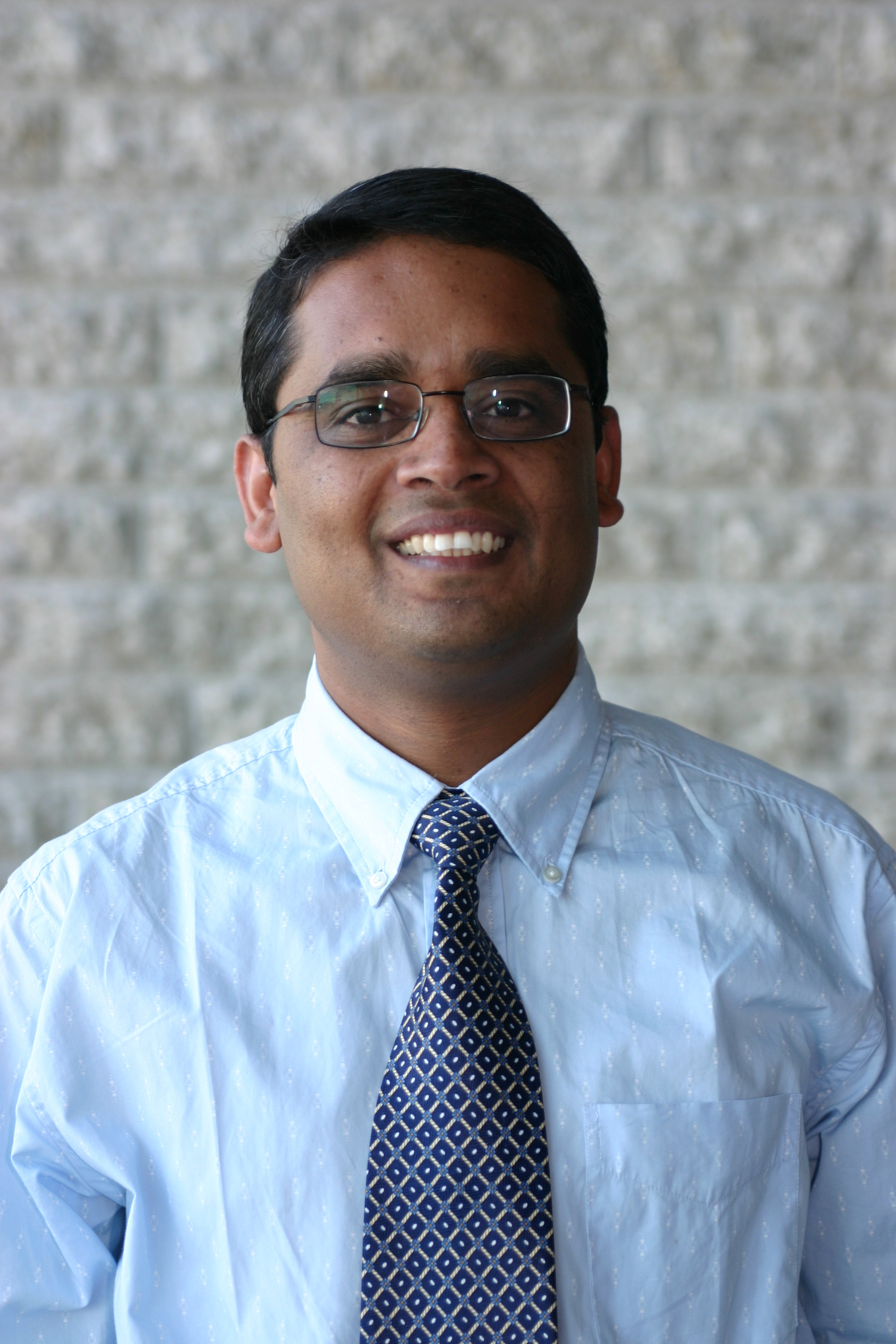}}]
{Ekram Hossain} (S'98-M'01-SM'06)  
is a Professor (since March 2010) in the Department of Electrical and Computer Engineering at University of Manitoba, Winnipeg, Canada. He received his Ph.D. in Electrical Engineering from University of Victoria, Canada, in 2001. Dr. Hossain's current research interests include design, analysis, and optimization of wireless/mobile communications networks, cognitive radio systems, and network economics.  He has authored/edited several books in these areas (http://home.cc.umanitoba.ca/$\sim$hossaina). His research has been widely cited in the literature (more than 7000 citations in
Google Scholar with an h-index of 42 until January 2014). Dr. Hossain  serves as the Editor-in-Chief for the {\em IEEE Communications Surveys and Tutorials}  and an Editor for {\em IEEE Journal on Selected Areas in Communications - Cognitive Radio Series} and {\em IEEE Wireless Communications}.  Also, he is a member of the IEEE Press Editorial Board. Previously, he served as the Area Editor for the {\em IEEE Transactions on Wireless Communications} in the area of  ``Resource Management and Multiple Access'' from 2009-2011 and an Editor for the IEEE Transactions on Mobile Computing from 2007-2012. He is also a member of the IEEE Press Editorial Board. Dr. Hossain has won several research awards including the University of Manitoba Merit Award in 2010 (for Research and Scholarly Activities), the 2011 IEEE Communications Society Fred Ellersick Prize Paper Award, and the IEEE Wireless Communications and Networking Conference 2012 (WCNC'12) Best Paper Award. He is a Distinguished Lecturer of the IEEE Communications Society (2012-2015). Dr. Hossain is a registered Professional Engineer in the province of Manitoba, Canada. 
\end{IEEEbiography}

\begin{IEEEbiography}[{\includegraphics[width=1in,height=1.25in,clip,keepaspectratio]{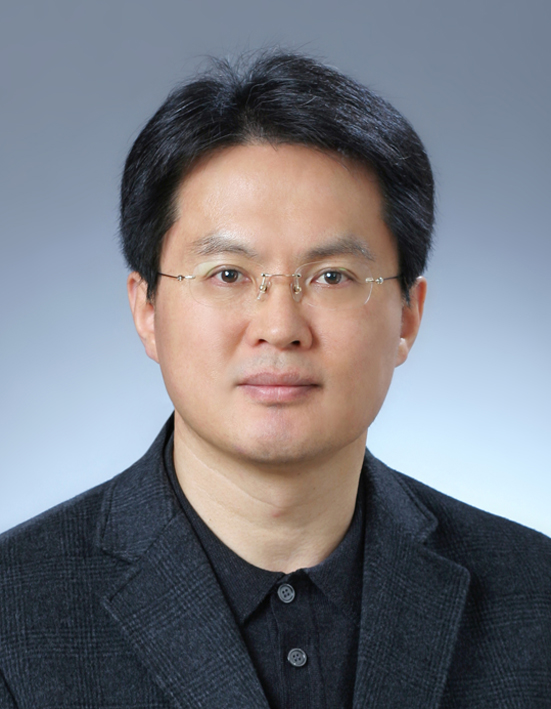}}]{Dong In Kim}
(S'89-M'91-SM'02) received the Ph.D. degree in electrical engineering from the University of Southern California, Los Angeles, in 1990. He was a tenured Professor with the School of Engineering Science, Simon Fraser University, Burnaby, British Columbia, Canada. Since 2007, he has been with Sungkyunkwan University (SKKU), Suwon, Korea, where he is currently a Professor with the College of Information and Communication Engineering. His research interests include wireless cellular, relay networks, and cross-layer design. Dr. Kim has served as an Editor and a Founding Area Editor of Cross-Layer Design and Optimization for the {\em IEEE Transactions on Wireless Communications} from 2002 to 2011. From 2008 to 2011, he served as the Co-Editor-in-Chief for the {\em Journal of Communications and Networks}. He is currently the Founding Editor-in-Chief for the {\em IEEE Wireless Communications Letters} and has been serving as an Editor of Spread Spectrum Transmission and Access for the {\em IEEE Transactions on Communications} since 2001.
\end{IEEEbiography}

\end{document}